\documentclass{ieeeaccess}
\usepackage{cite}
\usepackage{amsmath,amssymb,amsfonts}
\usepackage{graphicx}
\usepackage{textcomp}
\usepackage[
singlelinecheck=false 
]{caption}
\usepackage{lineno,hyperref}
\usepackage{subcaption}
\usepackage{fullpage}
\usepackage{times}
\usepackage{fancyhdr,graphicx,amsmath,amssymb}
\usepackage[ruled,vlined]{algorithm2e}
\usepackage{gensymb}
\usepackage[table,xcdraw]{xcolor}
\usepackage{array}
\newcommand\numberthis{\addtocounter{equation}{1}\tag{\theequation}}
\newcolumntype{P}[1]{>{\centering\arraybackslash}p{#1}}

\def\BibTeX{{\rm B\kern-.05em{\sc i\kern-.025em b}\kern-.08em
    T\kern-.1667em\lower.7ex\hbox{E}\kern-.125emX}}
\begin{document}
\history{Date of publication xxxx 00, 0000, date of current version xxxx 00, 0000.}
\doi{10.1109/ACCESS.2017.DOI}

\title{Combined Machine Learning and Physics-Based Forecaster for Intra-day and 1-Week Ahead Solar Irradiance Forecasting Under Variable Weather Conditions}

\author{\uppercase{Hugo Riggs}\authorrefmark{1}, \uppercase{Shahid Tufail}\authorrefmark{1},\uppercase{Mohd Tariq}\authorrefmark{1} \IEEEmembership{Senior Member, IEEE},
 and \uppercase{Arif Sarwat}\authorrefmark{1},\IEEEmembership{Senior Member, IEEE}}

\address[1]{Department of Electrical and Computer Engineering, Florida International University, Miami, FL 33174 USA}

\corresp{Corresponding authors:  Arif Sarwat (asarwat@fiu.edu) }

\begin{abstract}
Power systems engineers are actively developing larger power plants out of photovoltaics imposing some major challenges which include its intermittent power generation and its poor dispatchability. The issue is that PV is a variable generation source unless additional planning and system additions for mitigation of generation intermittencies. One underlying factor that can enhance the applications around mitigating distributed energy resource intermittency challenges is forecasting the generation output. This is challenging especially with renewable energy sources which are weather dependent as due to the random nature of weather variance. This work puts forth a forecasting model which uses the solar variables to produce a PV generation forecast and evaluates a set of machine learning models for this task. In this paper, a forecaster for irradiance prediction for intra-day is proposed. This forecaster is capable of forecasting 15 minutes and hourly irradiance up to one week ahead. The paper performed a correlation and sensitivity analysis of the strength of the relationship between local weather parameters and system generation. In this study performance of SVM, CART, ANN, and Ensemble learning were analyzed for the prediction of 15-minute intraday and day-ahead irradiance. The results show that SVM and Ensemble learning yielded the lowest MAE for 15-minute intraday and day-ahead irradiance, respectively.
\end{abstract}


\titlepgskip=-15pt

\maketitle

\section{Introduction}
\label{sec:introduction}
 \begin{figure*}
    \centering
    \includegraphics[height=8cm]{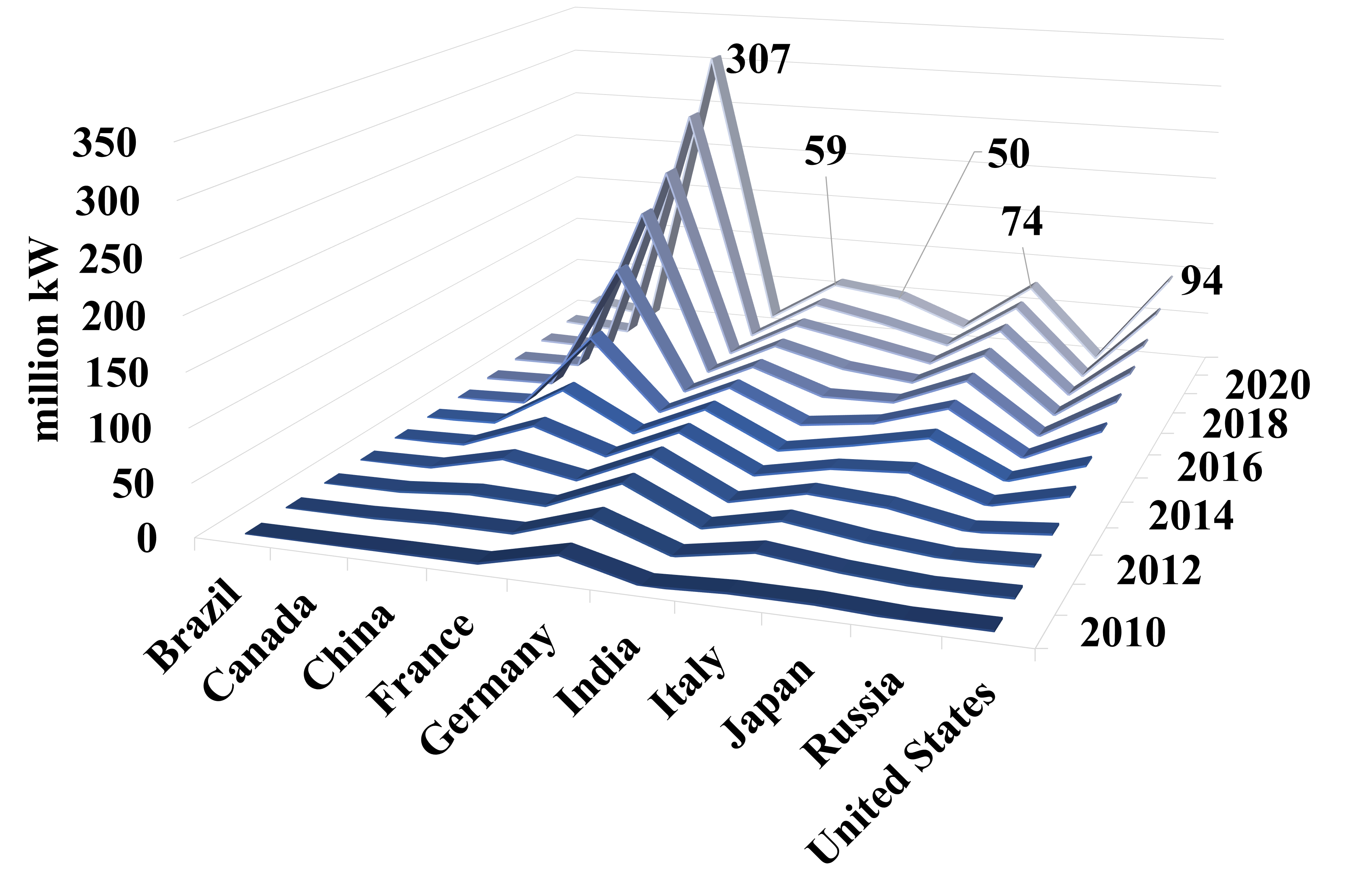}
    \caption{Total installed generation capacity of solar PV of the top ten countries from 2010 to 2020.}
    \label{fig:byCountryCapacity}
\end{figure*}
\begin{figure}
    \centering
    \includegraphics[width=\columnwidth]{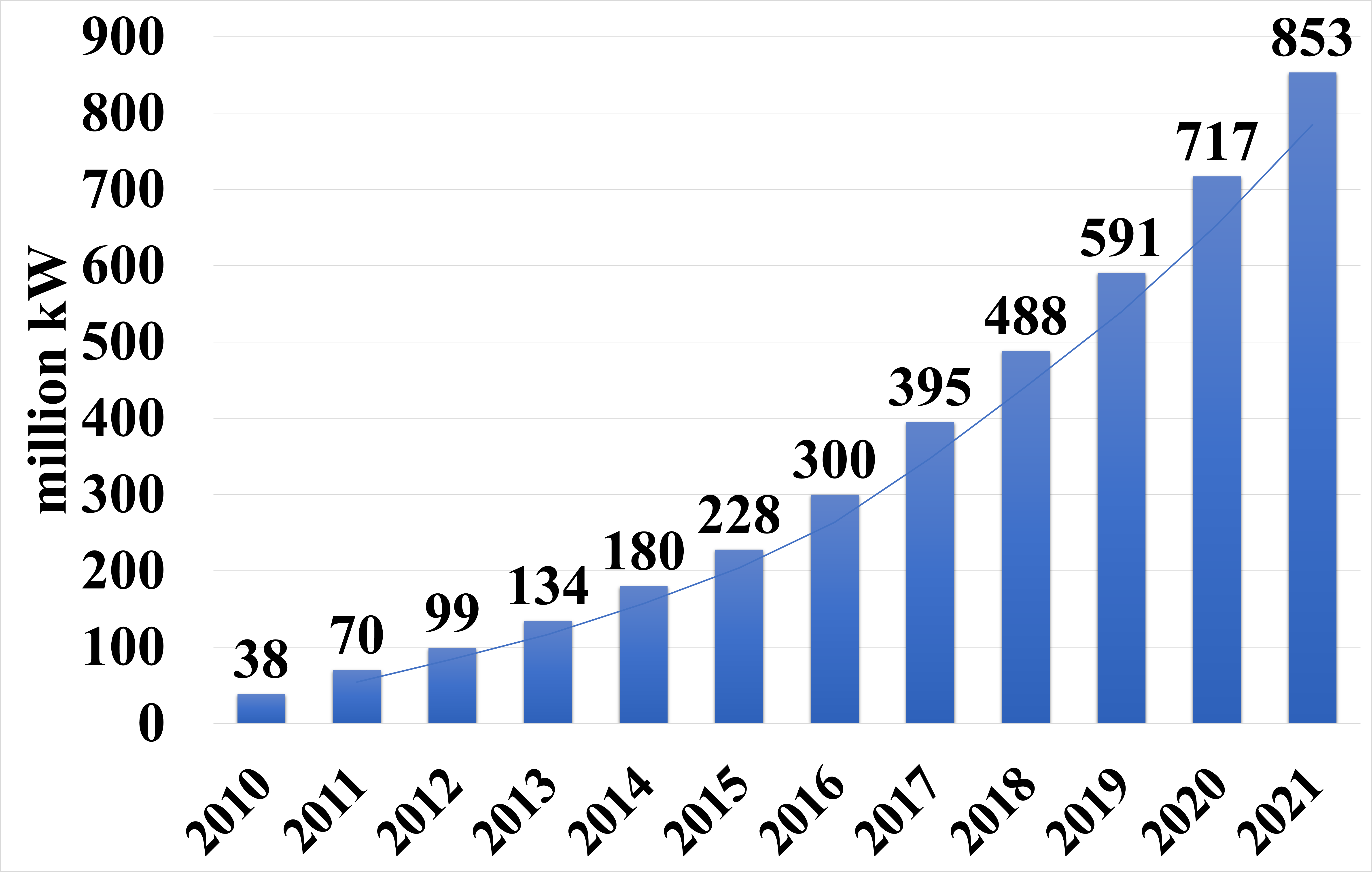}
    \caption{Total installed capacity of solar PV of the world from 2010 to 2021.}
    \label{fig:globalCapacity}
\end{figure}
Renewable energy sources have received considerable scientific attention as the depletion of fossil fuel resources accelerates. Global energy demand continues to increase through more industrialization and is driving the depletion of the oil and gas reserves \cite{holechek2022global, hussain2020impact, abbasi2021does}. The worlds energy consumption is from a wide range of generation sources that include renewable energy resources such as solar photovoltaic (PV) systems, wind-based energy systems, biomass-based generation, and tidal energy extraction schemes. 

As utility-scale PV Systems are being connected to power grids globally it is increasingly beneficial to know the future power output from these systems. The effects of local weather directly impact the power generation from PV systems causing intermittency of power  ~\cite{Aditya_2019_aggregation, Arif_2007}. Grid operators seek to enhance their visibility and control of their growing PV capacity and generation and recognize issues of providing a continuous power output from PV systems while reducing loss. The local weather has a high impact on PV generation consistency. The challenge of intermittent power output is addressed by a combination of technologies including energy storage management systems and forecasting technologies ~\cite{Arash_2016_combin, Sarwat2018}.  As PV penetration increases it has a greater potential to impact voltage and frequency fluctuations at the feeder level, and therefore advanced control and response mechanisms should be built for PV Systems that integrate the appropriate technologies to efficiently mitigate the intermittent nature of PV ~\cite{9764139, 9764108}.  

Especially over the past twelve years the installed capacity of solar energy has increased tremendously as countries bring PV into the fold of their generation mix. The top ten PV-adopting countries are expanding their PV generation capacity into the tens of gigawatts, as seen in figure \ref{fig:byCountryCapacity}. China has made up the majority of installed capacity having 307 GW of PV in the year 2020, and the United Sates is the second largest in installed capacity. The global capacity is on the trajectory to surprise one TW of PV generation in the early 2020's. The global increase in PV capacity is seen in figure \ref{fig:globalCapacity}.

Lorenz et al,  have modeled PV forecasting as a irradiance forecasting and subsequent PV estimation problem as well. The work implements a forecast across German PV canopies. Lerenz has also developed a confidence interval in forecasts. The confidence of the forecast will vary depending on dynamics with the weather and solar elevation. The confidence interval is defined by upper and lower limits. The standard deviations of error for hours of the day when GHI is above zero are calculated. The standard error in GHI is modeled over the total cloud coverage and the cosine of the solar zenith angle. The error distribution is determined by hourly forecasts. Forecast confidence upper and lower limits are calculated to be within two standard deviations around the expected forecast \cite{lorenz2009irradiance}.

The Forecast of variable generation by system operators paired with decreases in the time between dispatch schedules for generation can greatly increase access to flexible generation. Renewable forecasts can be applicable in grid functions of energy storage and dispatch increasing the flexibility of the grid ~\cite{mills_dark_2011}. System operators can increase system flexibility through the dynamic generation and load balancing mechanisms with high solar penetration levels, and they desire high-resolution and low error predictions to reduce cost on utilities to implement.  

The approach proposed in this work is an in-direct forecast of PV power by forecasting the irradiance and computing the expected power and energy from the irradiance forecast. The estimator utilized has very low error in estimating PV power given irradiance and module temperature. A model of solar position is implemented for calculating the ideal irradiance. Four regression models are developed based on local weather and ideal irradiance. The models implemented for irradiance regression are the artificial neural network (ANN), the support vector machine (SVM) for regression, the classification and regression tree (CART), and the ensemble forecaster. A set of studies for the forecast of PV production is created based on real PV systems in Florida. The key contributions of this paper are that it: 

\begin{enumerate} 
    \item Presents a forecaster for Irradiance prediction for intra-day, (i.e. 15 minutes and hourly) up to one week ahead. 
    \item Provides system model description, methodologies, and diagrams, along with correlation and sensitivity analysis of the strength of the relationship between local weather parameters and system generation.
    \item Estimates production from a 1.4 MW DC PV canopy site in Florida using irradiance forecasting, and details the error of the forecast for short-term and medium-term forcasts.
\end{enumerate}

\subsection{Related Work}
The techniques to estimate PV generation have been applied in the literature, and PV system performance has been well described in the literature ~\cite{Aditya_2019_aggregation, Aditya_2019_CaseStudy}. The breadth of this paper is to demonstrate the forecasting capabilities and approach. In ~\cite{Olowu_2019} the aggregation of PV systems shows reduced variability in the production, where the larger the aggregated system the more the variability is decreased.  

Paatero et al. have shown ~\cite{PARIDA20111625} that voltage problems in middle voltage distribution networks are likely as PV penetration increases, which will involve voltage drop at feeder levels which can be damaging to hardware and grid stability. In ~\cite{Aditya_2019_CaseStudy} a data-driven estimator is used to convert irradiance to KW and KWH values. 

A comprehensive review of irradiance forecasting research is provided in ~\cite{DIAGNE201365}. The ARIMA model is used in ~\cite{reikard_predicting_2009} along with comparisons with neural network model for intraday forecasts. They report errors up to 39\% and it varies largely. In ~\cite{perez_validation_2010}, the use of satellite images for intraday forecasts is shown. They conclude that clear-day irradiance accuracy is very high and cloudy-day irradiance forecasts have much higher inaccuracy. In \cite{Lorenz} two approaches are used, a strictly numerical weather prediction forecaster and one based on images either from sky-imaging or from the satellite. The work targets intra-day forecasting and shows that both techniques outperform the persistence forecast.   

Forecasting of a mid-level household has shown the ability to model weather and observed generation into an online learning model for accurate PV generation prediction ~\cite{bashir2019solar}. However, this model relies on regional weather forecasting and does not implement more local weather analysis. This paper stands out by presenting forecasting from irradiance to KW and KWH using known system DC capacity. 
\begin{figure*}[h]
    \centering
    \includegraphics[width=15cm]{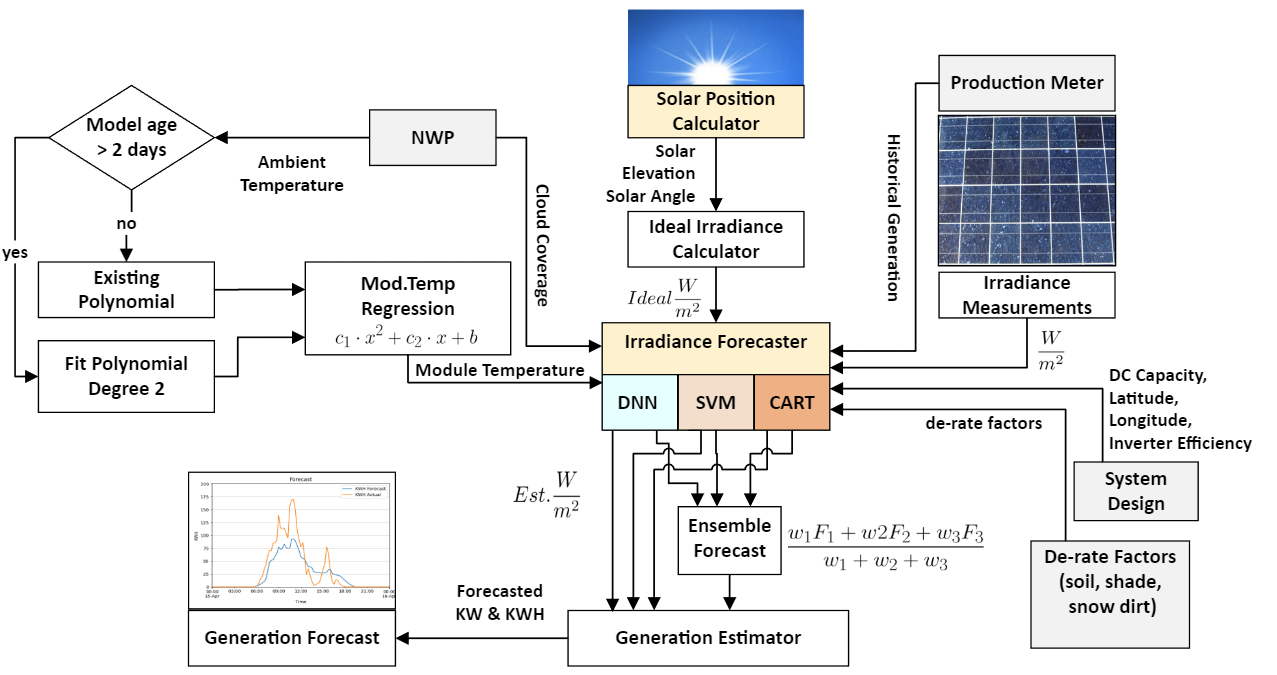}
    \caption{The system model diagram of the combined forecaster.}
    \label{fig:systemmodel}
\end{figure*}
This paper distinguishes itself from previous related works in developing an irradiance forecaster for PV generation forecasts that implements various sequential regression models for processing the final forecast. Our model includes derate factors, which are environmental elements and system design parameters that influence the AC power generated by a PV system [37]. These parameters include not just the DC-to-AC conversion efficiency of the string inverter but also losses from DC and AC wiring, soiling, shading due to adjacent PV modules, trees and buildings, and module mismatch. In Table 1, these parameters are given as scalar values, and the product of these values reflects the overall derating factor for the PV system. Therefore, this presents the need for forecasting methods that account for losses and derate factors and metrics that can compare performances of differently sized PV systems located at regions experiencing different weather conditions. 
\begin{table}[]
\centering
\captionsetup{justification=centering}
    \captionof{table}{The parameters of the PV system.} 
\begin{tabular}{lc}
\hline
\rowcolor[HTML]{EFEFEF} 
\textit{\textbf{Parameter}}            & \textit{\textbf{PV Array}}              \\[2pt] \hline
location                      & Miami                               \\[2pt]
latitude-longitude            & 25.76$\degree$N,                    \\[2pt]
                              & \multicolumn{1}{l}{80.36$\degree$W} \\[2pt]
elevation, ft                 & 10                                  \\[2pt]
nameplate rating, MW DC       & 1.4                                 \\[2pt]
AC capacity, MW               & 1.104                               \\[2pt]
number of inverters           & 46                                  \\[2pt]
inverter efficiency  \%         & 98                                \\[2pt]
number of PV modules          & 4480                                \\[2pt]
module efficiency, \%         & 16.5                                \\[2pt]
number of strings in series x & 56 x 4                              \\[2pt]
number of arrays              & \multicolumn{1}{l}{}                \\[2pt]
modules per string            & 20                                  \\[2pt]
tilt, azimuth of array        & 5$\degree$, 2678$\degree$           \\[2pt]
soiling derate factor         & 0.9                                 \\[2pt]
cabling loss factor           & 0.99                                \\[2pt]
temperature coefficient       & -0.5                                \\[2pt]
module mismatch factor        & 0.97                                \\ \hline
\end{tabular}
\label{tab:parameters}
\end{table}
The rest of the paper is as follows. Section \ref{sec:model} provides the system model for the solar position, ideal irradiance, solar generation the forecaster, and sensitivity and correlation analysis of the solar variables. Section \ref{sec:ml} presents the machine learning models. Section \ref{sec:results} provides the studies and results of various and distinct studies for the forecaster. Section \ref{sec:conclusion} concludes the paper.   
\section{System Models and Framework of Methodology}
\label{sec:model}
The forecaster involves a solar position calculator that calculates the solar elevation and angle. The angle and elevation of the sun are used to calculate the ideal irradiance. The ideal irradiance is calculated over a given time range and is used by the irradiance forecaster. Historical generation data from the production meter is passed to the irradiance forecaster. Numerical weather prediction (NWP) of the next seven days is captured and used for the module temperature regression model, which uses the NWP ambient temperature to predict module temperature subsequently, module temperature is input for the irradiance forecaster. The irradiance forecaster implements three machine-learning models for irradiance forecasting from the input data. The ensemble forecast averages these models' forecasts. The generation estimator provides the generation forecast from the irradiance forecast. The system is shown in Fig \ref{fig:systemmodel}.  The temperature time series of module temperature is referred to as {Mod.Temp} follows the ambient temperature with a slight time lag and is well fit by a 2-degree polynomial regression model in our forecaster. If the module temperature is not available, that data can be generated by the  forecaster from NWP ambient temperature.  At a standard temperature of 25\degree C the panels perform well; for every degree above the standard, the efficiency of the panels will drop by a small percentage. This follows that for each degree below the standard, the efficiency of the panels will increase. Specific changes in efficiency do vary by the manufacturer and model of the PV panel ~\cite{654300}.  The system works through an online  training algorithm that updates regression models on a rolling window. It has been executed on 4 years of data from 2017 to 2021. When the model retention window is expired a new model is trained, this is currently set for daily and can easily train overnight before the next day.

\subsection{Quantatative Data Analysis}
\begin{figure*}[t]
    \centering
    \captionsetup{justification=centering}
    \includegraphics[width=16.1cm]{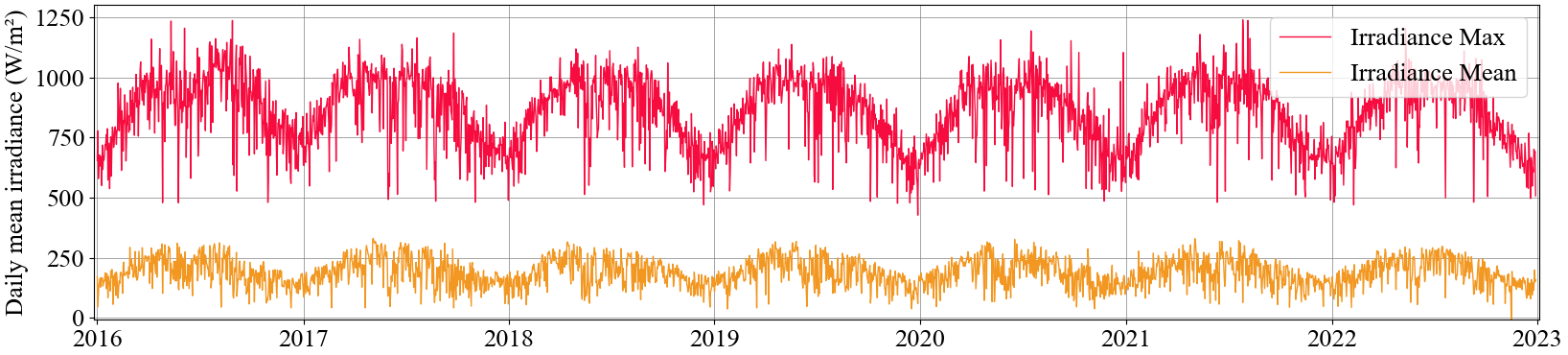}
    \caption{The time series plot of irradiance recorded from the College of Engineering and Computing, in Miami.}
    \label{fig:historical}
\end{figure*}

Data sets composed of regularly sampled meteorological variables linked to solar energy from two different Florida locations of Miami and Daytona were analyzed. These variables include irradiance, module temperature, ambient temperature, and cloud coverage. The data's quantitative study yielded the maximum, minimum, median, mean, and standard deviation values. The third-standard deviation and fourth-standard deviation are removed as outliers.  Data imputation, sampling averages from available years, in applied forward and back-filling is performed to fill gaps in data. The clean and imputed gap-filled data for irradiance on the Miami location is shown in figure \ref{fig:historical}. This prepares the data for the supervised training at the task of short-term and medium-term irradiance prediction.
\begin{figure}
    \includegraphics[width=1\columnwidth,trim={0.2cm 0.2cm 0.2cm 1cm},clip]{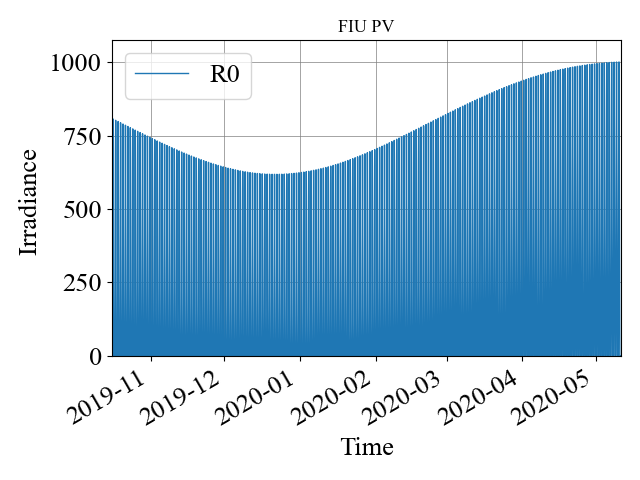}
    \caption{A portion of the output from the clear sky model of global horizontal irradiance.}
    \label{fig:my_label}
\end{figure}
\begin{figure*}[ht]
    \centering
    \begin{subfigure}{.248\textwidth}
      \centering
      \includegraphics[width=1\linewidth]{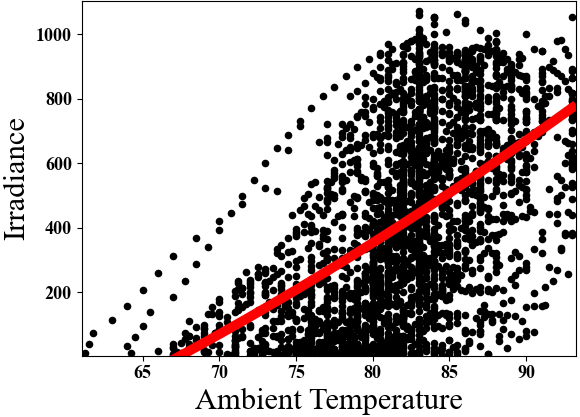}
      \label{fig:sub2}
    \end{subfigure}
    \begin{subfigure}{.248\textwidth}
      \centering
      \includegraphics[width=1\linewidth]{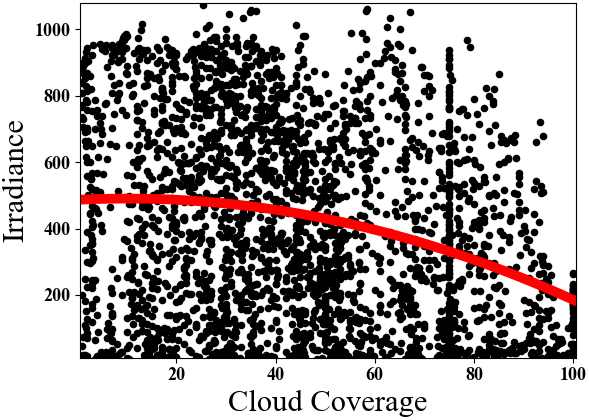}
      \label{fig:sub1}
    \end{subfigure}%
    \label{fig:test}%
    \centering
    \begin{subfigure}{.248\textwidth}
      \centering
      \includegraphics[width=1\linewidth]{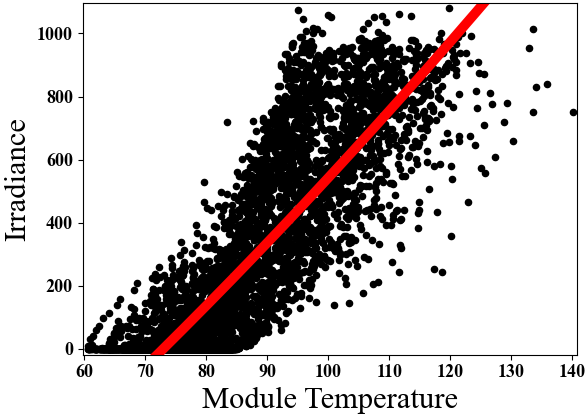}
      \label{fig:sub3}
    \end{subfigure}
    \begin{subfigure}{.248\textwidth}
      \centering
      \includegraphics[width=1\linewidth]{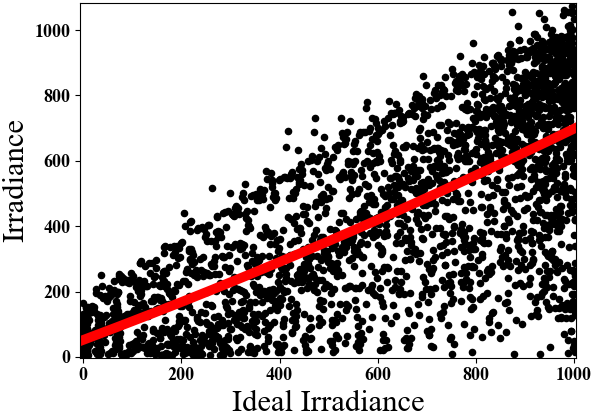}
      \label{fig:sub1}
    \end{subfigure}%
    \caption{Polynomial linear regression functions fit to the different variables in the data set.}
\end{figure*}
\subsection{Model of Solar Position}\label{sec:Forecaster}
While the earth travels around the sun, for the purpose of tracking the solar position the sun is considered as the moving component in relation to a fixed position on earth in a spherical coordinate system. Solar radiation travels towards the earth\textquotesingle s surface and this radiation is absorbed by parts of the atmosphere, which results in an irregular spectrum reaching the surface of the earth. The natural solar resource is modeled based on the following set of equations, which assume the observer is in the northern hemisphere. 

Due to daylight savings times, and different time zones the true solar noon, a time when the sun is at its peak elevation, will differ from the local noon time. The hour angle of the sun is calculated based on the following. 
\begin{equation} 
    H = \left ( \frac{15\degree}{hour} \right ) \cdot (\text{hours before solar noon}) 
\end{equation}
\textbf{Julian Calendar:} the Julian calendar is used by astronomers for facilitating calculations of solar position. A Julian date is represented in computation as a single number and thus it facilitates calculations and comparisons of time. The Julian date is calculated from the Gregorian date by the following equation:
\begin{align*}
    \text{JD} &= \left \lfloor 365.25 \times year + 4716 \right \rfloor \\ 
    & + \left \lfloor 30.6001 \times (month + 1)  \right \rfloor \\
    & + day + B - 1524.5
\end{align*}
Where $year$, $month$, and $day$ are the Gregorian calendar values, and $B$ is calculated from the Gregorian year as follows.
\begin{equation} 
    B = 2 - \left \lfloor \frac{year}{100} \right \rfloor + \left \lfloor \frac{year}{400} \right \rfloor
\end{equation}
Solar declination is the angle of the sun from the equator, which will peak at the summer solstice and reach its lowest value at the winter solstice, the following equation is used:
\begin{equation} 
     \delta = 23.45\sin[\frac{360}{365}(n-81)] 
\end{equation} 
Where $\delta$ is the solar declination. The value $n$ is the day of the year such that $n=1$ is January 1\textsuperscript{st}, $n=182$ is July 1\textsuperscript{st}, and $n=365$ will be December 31\textsuperscript{st}.
Azimuth is the angle of the line passing through due north and the line passing through the sun. 

To determine the altitude angle the following equation is used:
\begin{equation} 
   \beta = \sin^{-1} \left ( \cos L \cos \delta \cos H + \sin L \sin \delta \right )
\end{equation} 
Where $\beta$ is the altitude angle of the sun, $L$ is the latitude on earth, $H$ is the hour angle of the sun and $\delta$ is the solar declination.
\begin{equation} 
    \sin \phi s = \frac{\cos \delta \sin H}{\cos \beta} 
\end{equation}
As the inverse of sine is ambiguous, $\sin x = \sin(180 - x)$, a test is applied to determine if the azimuth is greater than or less than $90\degree$ away from south.
\begin{align*}
    &\text{if } \cos H \ge \frac{\tan \delta}{\tan L}\text{, then }\\
    &|\phi_{S}| \le 90\degree \text{;} \\
    & \text{ otherwise    } |\phi_{S}| > 90\degree
    \numberthis \label{eqn}
\end{align*}
Where $\phi s$ is the sun's azimuth, within summer months the azimuth may exceed $90\degree$ away from the south.
The ideal GHI is calculated using the average solar altitude angle within the following polynomial. 
\begin{align*}
    GHI &= -4.72^{-7}\left [  \frac{\beta_i+\beta_{i+1}}{2}\right ]^{5}+ 1.15^{-4}\left [  \frac{\beta_i+\beta_{i+1}}{2}\right ]^{4}\\ 
    &- 1.15^{2}\left [  \frac{\beta_i+\beta_{i+1}}{2}\right ]^{3}+ 4.78^{-1}\left [  \frac{\beta_i+\beta_{i+1}}{2}\right ]^{2} \\
    &+ 8.31\left [  \frac{\beta_i+\beta_{i+1}}{2}\right ] - 0.079
\end{align*}
Such that $\beta_i$ is the altitude angle at a date-time index. Ideal GHI irradiance is modeled with a five-degree polynomial fit to the clear-day irradiance utilizing the average altitude angle $\left [  \frac{\beta_i+\beta_{i+1}}{2}\right ]$ as the independent variable.
\subsection{PV System and Generation Estimation}
\begin{figure*}
   \centering
    \captionsetup{justification=centering}
    \includegraphics[width=16.1cm]{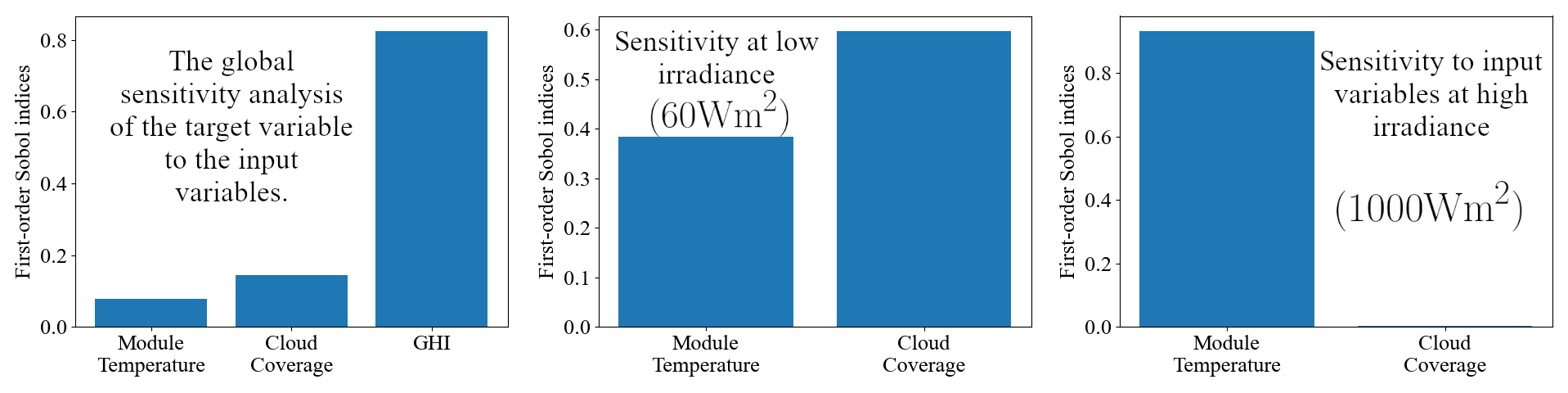}
    \caption{The Sobol indices, sensitivity, values from Monte Carlo estimation of the ANN.}
    \label{fig:sensitivity}
\end{figure*}
{\renewcommand{\arraystretch}{1.4}
\begin{table}[b]
    \caption{Correlation of the solar variables with generation.}
    \label{tab:cor}
\centering
\begin{tabular}{p{0.45\columnwidth}|cp{0.45\columnwidth}}
\hline
\rowcolor[HTML]{EFEFEF} 
\textit{\textbf{Solar Variables}} & \textit{\textbf{Correlation}} \\ \hline
Ideal Irradiance                  & 0.74                          \\ \hline
Irradiance                        & 0.98                          \\ \hline
Ambient Temperature               & 0.53                          \\ \hline
Module Temperature                & 0.68                          \\ \hline
Cloud Coverage                  & -0.2                          \\ \hline
\end{tabular}
\end{table}}
The model of PV estimation is comprehensive, using different system design characteristics and environmental derate factors that affect the AC power generated by the PV system. These factors include DC to AC conversion efficiency of the string inverters, the losses caused by DC and AC cabling, the buildup of particles on the surface of the PV (soiling), and shadows cast over the system's modules by other objects such as buildings (shading). These derate factors are shown in table \ref{tab:parameters}.
\begin{equation} \label{eq:estimate}
        \begin{split} 
        KW AC_{estimated} = KW DC_{rated}\frac{R_{i}}{1000}[1 
            \\+\frac{Temp_{Coef}}{100}\{{T_{i} - T_{module_{avg}})}\}]P 
        \end{split} 
\end{equation} 
Where $P$ are the combined derate factors and $R_{i}$ and $T_{i}$ are the irradiance and module temperature at moment $i$ in the forecast cycle respectively.
The generation estimator, used in \cite{8336582}, is applied and requires the irradiance as an input in calculating the power output. In the following section, three machine learning irradiance regression models and a fourth ensemble model are presented which calculate the expected irradiance using the solar position, module temperature, and cloud coverage.  

Derate factors are considered as cable loss, mismatch, inverter efficiency, and temperature  coefficient~\cite{Aditya_2019_aggregation}. A PV system power output curve is similar to the local irradiance curve, ideal irradiance can be calculated based on the solar position which consists of two parameters, azimuth and elevation for a specific time of the day. A PV system power output curve is similar to the local irradiance curve and also affected by ambient temperature. The cloud coverage over solar panels reduces generation, also the module temperature of the panels has a strong effect on their efficiency. 

Where $R0$ is ideal clear sky irradiance and $n$ is the cloud coverage percentage. 
\begin{figure*}[t]
    \centering
    \captionsetup{justification=centering}
    \includegraphics[width=16.1cm,height=10cm]{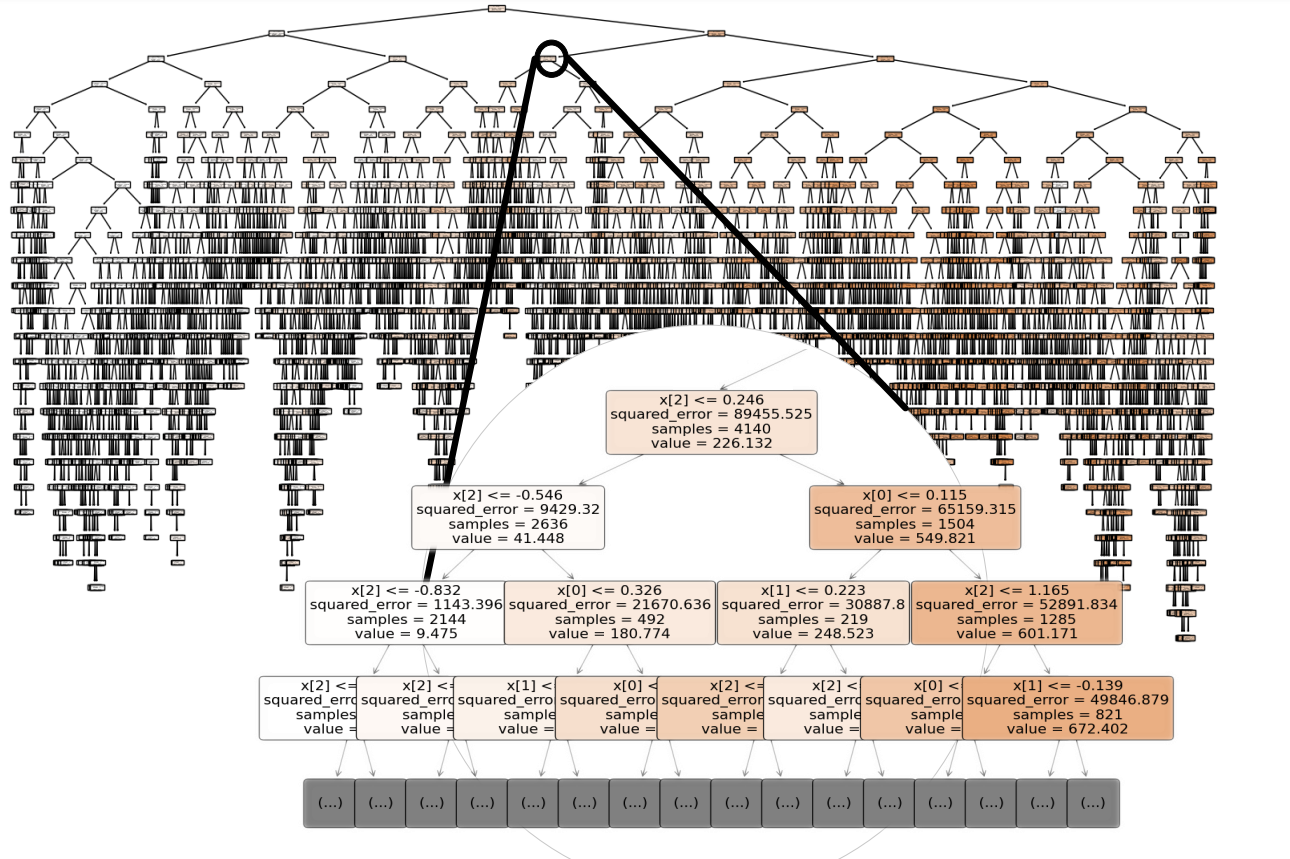}
    \caption{Classification and Regression Tree After Training on Solar Data}
    \label{fig:cart}
\end{figure*}
\subsection{Correlation and Sensitivity  Analysis}
The correlation is computed based on the pearson correlation coefficient.
\noindent\textbf{Pearson correlation:} The Pearson product-moment correlation coefficient, $r_{xy}$, is used to analyze the relationships between weather and irradiance and generation. The correlation coefficient is calculated by:
\begin{equation}
\label{eq:corr}
  r _{xy} = \frac{\sum _{i=m}^p(x_i -\bar x)(y_i -\bar y))}{\sqrt{\sum _{i=m}^p(x_i -\bar x)^2}\sqrt{\sum _{i=m}^p(y_i -\bar y)^2}}  
\end{equation}
where the weather and generation data at historical time $m$ is calculated until the present time $p$. The correlation is calculated for $x_i$ and $y_i$ over the data set, where $\bar x$ and $\bar y$ are the means, respectively. The correlation coefficients for all weather parameters and generation are summarized in table \ref{tab:cor}. The correlation between irradiance and module temperature to the generation is high at greater than 86\%. There is a low negative correlation between cloud coverage and irradiance across the data. analysis shows strong correlations between irradiance, solar elevation, and also that module temperature is more closely correlated with generation than ambient temperature alone. The correlation of the solar variables with generation are summarized in table \ref{tab:cor}. 
The sensitivity analysis shows that generation, the $kW$ variable, is highly sensitive to irradiance, and also significantly sensitive to module temperature, and ambient temperature to a lesser degree. 
\begin{align*}
    \text{V}_{X_{i}}(E_{X_{ \sim i}}(Y|X_i)) &\approx \\
    &\frac{1}{N}\sum_{j=1}^{N}f(\textbf{B})_j \left ( f(\textbf{A}^{i}_{B})_j - f(\textbf{A}_j)\right )
\end{align*}
Sensitivity analysis of the solar variables is performed by decomposition of the output variance using the Sobol indices with a Monte Carlo estimator. It is found that the ideal GHI input variable has the most effect on the target variable. It is also found that cloud coverage has a higher impact on the target when the irradiance is fixed at a lower value which suggests that cloud coverage can have a stronger effect on expected irradiance around the hours of sunrise and sunset. The Sobol indices are calculated from the given equation from Sobol et.al, \cite{sobol2001global,saltelli2010variance}. In which $\textbf{A}$ and $\textbf{B}$  represent samples of input matrices. The estimator activates the models predict function $f()$ and models the variance for a $j^{th}$ trial. Variance $V_{X_{i}}(\cdot )E_{X_{i}}(\cdot )$ or mean of argument $(\cdot )$ taken over all factors. The notation $A^{(i)}_B$ is the matrix, where column $i$ comes from matrix $B$ and all other columns come from matrix $A$.
\section{Machine Learning Models}
\label{sec:ml}
\begin{figure}
    \centering
    \includegraphics[width=1\columnwidth]{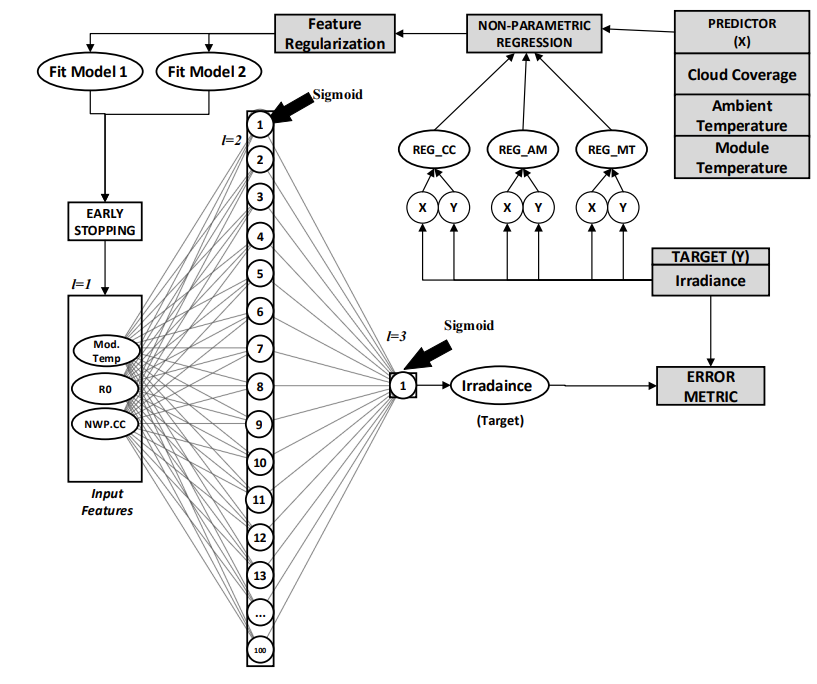}
    \caption{Framework of the artificial neural network learning process.}
    \label{fig:DNN}
\end{figure}
\subsection{Classification and Regression Tree}
The classification and regression tree (CART) model is a decision tree learning process that creates branching decisions on the input feature domain towards many leaf nodes that encompass the output range of the function. A decision tree itself can be used to extract the pattern from data and predict the target given the input features. Individual internal nodes in the decision tree are referred to as input features, and each leaf node of the tree is the best approximate value to expect through the regression model based on a historical fit of the data. The algorithm for the CART is applied and the fully formed tree structure, after fitting the data, is shown in figure \ref{fig:cart}. This is a supervised learning  algorithm that uses variance as the cost function and attempts to minimize variance during the selection of node splits while building the tree. 

\subsection{Artificial Neural Network}
The artificial neural network (ANN) is a seven-layer feedforward network design. This model utilizes the Sigmoid activation function as shown in equation \ref{eq:sigmoid}. 
\begin{equation}\label{eq:weight}
    x = \sum_{i=1}^{n} x_{i}w_{i}
\end{equation}
The input for the Sigmoid activation is calculated from the weight and bias data of each incoming edge connection. 
\begin{equation}\label{eq:sigmoid}
    \sigma(x) = \frac{1}{1+e^{-x}}
\end{equation}
This weight and bias information is summed across input edges to the node and provided as input to the activation function. This weight information sum is shown in equation \ref{eq:weight}.  Input forecasted data for ideal irradiance $R0$ cloud coverage $CC$ and module temperature in Fahrenheit $Mod.Temp$ are fed to the model. Using a supervised machine learning process with backpropagation of the error, the ANN learns to predict the target variable of irradiance $R$. The input data has been scaled, with the min-max scaling technique, from training data, applying the same scaler object to testing data. The input data for module temperature is retrieved from the regression model to provide the predicted module temperature to be used as input for the irradiance prediction. The early stopping callback function is used to save the ANN model's weights upon reaching a minimum validation loss during training. The criteria for stopping is that two epochs pass without decreasing the training loss at a 0.01 tolerance minimum.  
\begin{figure*}[h]
    \captionsetup[subfigure]{justification=centering}
    \begin{subfigure}{\columnwidth}
        \centering
        \includegraphics[width=\columnwidth,trim={0 2cm 0 1cm},clip]{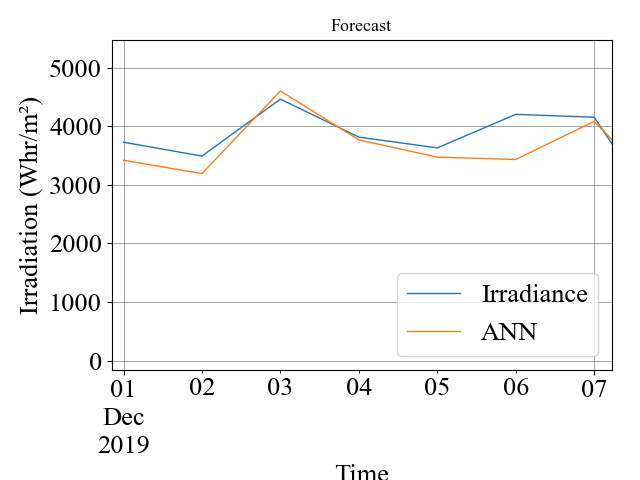}
        \caption{ANN irradiation forecast.}
        \label{fig:ann_week_irradiation_daily}
    \end{subfigure} 
    \begin{subfigure}{\columnwidth}
        \centering
        \includegraphics[width=\columnwidth,trim={0 2cm 0 1cm},clip]{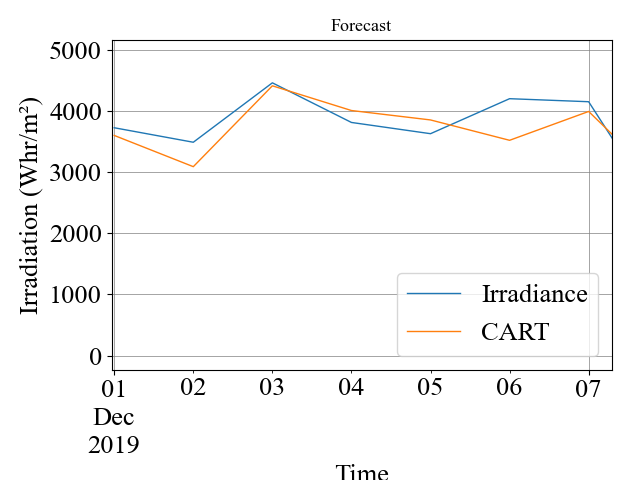}
        \caption{CART irradiation forecast.}
        \label{fig:cart_week_irradiation_daily}
    \end{subfigure}
    \begin{subfigure}{\columnwidth}
        \centering
        \includegraphics[width=\columnwidth,trim={0 2cm 0 1cm},clip]{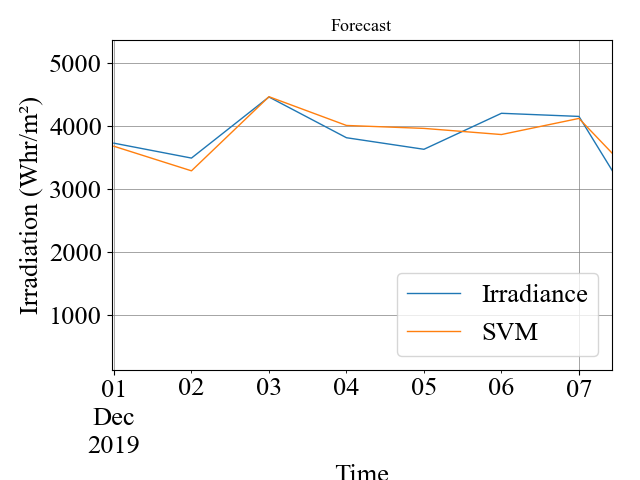}
        \caption{SVM irradiation forecast.}
        \label{fig:svm_week_irradiation_daily}
    \end{subfigure}
    \hspace*{0.55cm}
    \begin{subfigure}{\columnwidth}
        \centering
        \includegraphics[width=\columnwidth,trim={0 2cm 0 1cm},clip]{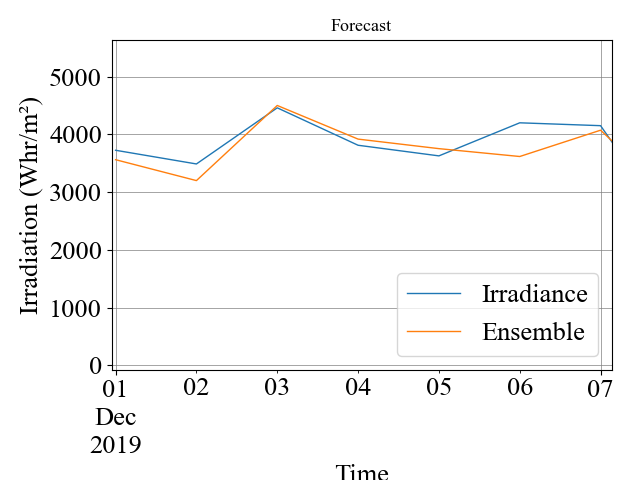}
        \caption{Ensemble irradiation forecast.}
        \label{fig:ensemble_week_irradiation_daily}
    \end{subfigure}
    \caption{The forecast of one week of irradiation $Whr/m^2$}
\end{figure*}
\subsection{Support Vector Machine}
A support vector machine (SVM) is fit to the training data to function as a regression model. The SVM performs an optimization considering the set of samples $\left \{ (x_{1},y_{1}),...,(x_{k},y_{k})  \right \}$, where $x_{i} \in R^{n}$ is the feature vector, $y_i$ is the target output, and $k$ is the number of samples in the training set. The standard epsilon support vector regression form is used \cite{chang2011libsvm}. The parameters are $C=1000$, $\epsilon=0.1$. 
\begin{align*}
    &\min_{w,b,\xi,\xi^{*}} &\frac{1}{2}w^{T}w+C\sum_{i=1}^{k}\xi_{i}+C\sum_{i=1}^{k}\xi^{*}_{i}\\
    &\text{subject to } &w^{T}\phi(x_{i})+b-y_{i}\leq \epsilon + \xi_{i},\\
    &&y_{i}-w^{T}\phi(x_{i})-b\leq \epsilon + \xi_{i}^{*},\\
    &&\xi_{i},\xi_{i}^{*} \geq 0, i=1,...,k
\end{align*}
The support vector weight information of $w$ is computed and cached for each feature in $x_{i}$ that is within the margin boundary. The radial basis function kernel is selected. In its operation, the SVM calculates and caches the support vectors which minimize the distance between the function output and the target variable. The $\xi$ variables define the $\xi$ in-sensitive loss, and the kernel will capture most of the data in optimizing the model fit but will not count loss from variables outside the $\xi$ defined region. Each model's efficacy was validated by comparison of the results with the values measured at the solar canopy. Based on the analysis of the prediction models'  results, an ensemble model which does weight averaging of the other model's outputs, in generating the ensemble forecast, is the best prediction of irradiance. In section \ref{sec:results} it is shown that the ensemble model's output lag plot is most similar to the actual irradiance lag plot in comparison with the other models. 

Each output value from the machine learning models is averaged in the ensemble forecast. The ensemble forecast uses the relative root mean squared error. The inverse of the RRMSE is used to weight each forecast input into the ensemble model. The ensemble model averages the weighted inputs over the sum of the inverse RRMSE values. The RRMSE relates the mean error to the predicted values as seen in equation \ref{eq:rrmse}
\begin{equation}\label{eq:rrmse}
RRMSE = \sqrt{\frac{\frac{1}{N}\sum_{i=0}^N(y_i - \hat y_i)^2}{\sum_{i=0}^N(\hat y_i)^2}}
\end{equation}
The ensemble model uses this information in the form of the equation \ref{eq:ensemble} to calculate its output.
\begin{equation}\label{eq:ensemble}
    \text{ensemble} = \frac{w_1F_{1}+w2F_2+w_3F_3}{w_1+w_2+w_3}
\end{equation}
In this formulation, the input forecasts from the DNN, SVM, and ANN models are taken as $F_{1}$, $F_2$, and $F_3$ and the weights are the inverse RRMSE scores for each model $w_1$, $w_2$, and $w_3$ calculated as $w_1 = DNN_{RRMSE}^{-1}$, the value $w_2 = SVM_{RRMSE}^{-1}$, and $w_3 = CART_{RRMSE}^{-1}$; thus a reduced relative error of the model leads to a  greater contribution to the ensemble model's output.
\begin{figure}
    \centering
    \includegraphics[width=\columnwidth]{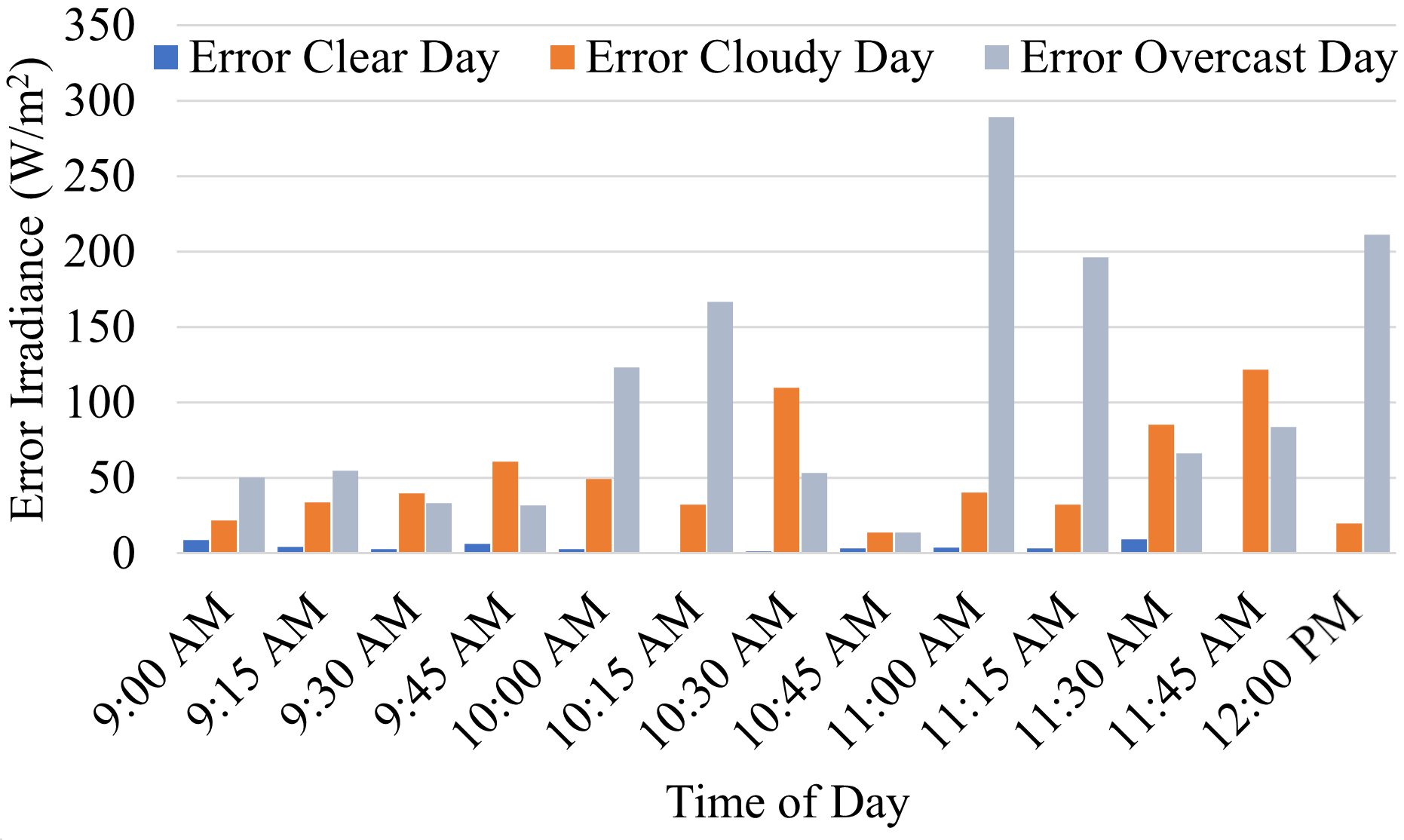}
    \caption{The absolute error of forecasts on mornings of days selected by cloud coverage amount.}
    \label{fig:clouderror}
\end{figure}
\begin{figure*}[h]
    \captionsetup[subfigure]{labelformat=empty,justification=centering}
    \begin{subfigure}{\columnwidth}
        \centering
        \includegraphics[width=\columnwidth,trim={0 1.2cm 0 1cm},clip]{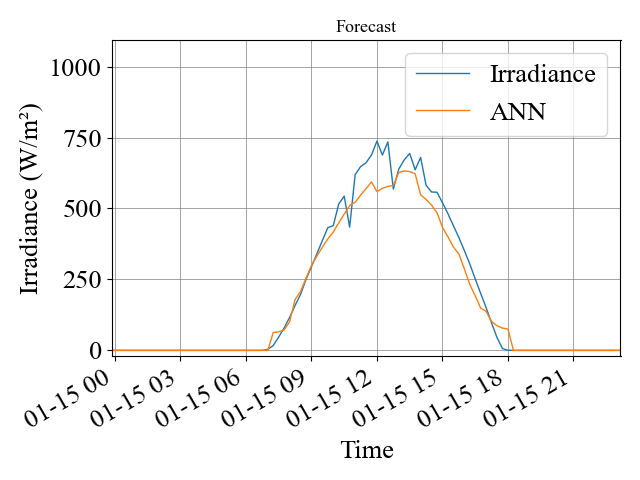}
        \caption{ANN irradiance forecast for January $15^{th}$.}
        \label{fig:ann_jan}
    \end{subfigure} 
    \begin{subfigure}{\columnwidth}
        \centering
        \includegraphics[width=\columnwidth,trim={0 1.2cm 0 1cm},clip]{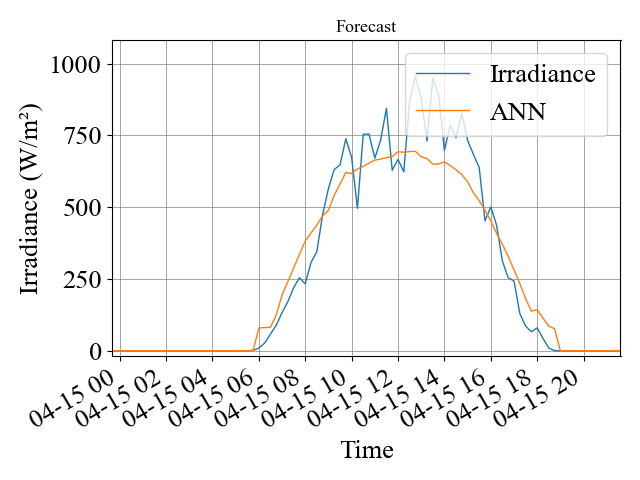}
        \caption{ANN irradiance forecast for April $15^{th}$.}
        \label{fig:ann_apr}
    \end{subfigure}
    \begin{subfigure}{\columnwidth}
        \centering
        \includegraphics[width=\columnwidth,trim={0 1.2cm 0 1cm},clip]{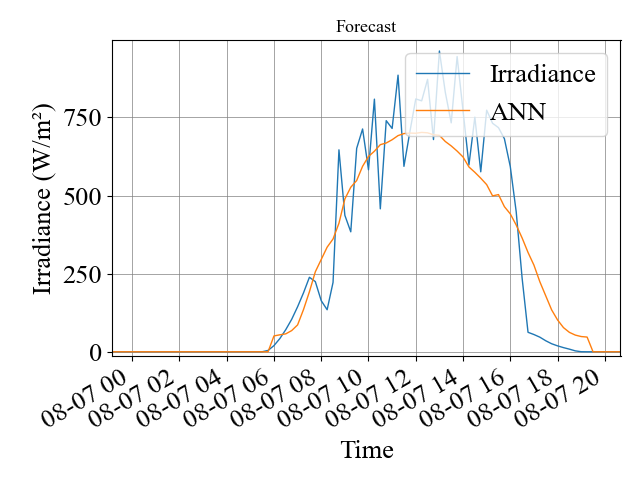}
        \caption{ANN irradiance forecast for August $7^{th}$.}
        \label{fig:ann_aug}
        \end{subfigure}
    \hspace*{0.55cm}
    \begin{subfigure}{\columnwidth}
        \centering
        \includegraphics[width=\columnwidth,trim={0 1.2cm 0 1cm},clip]{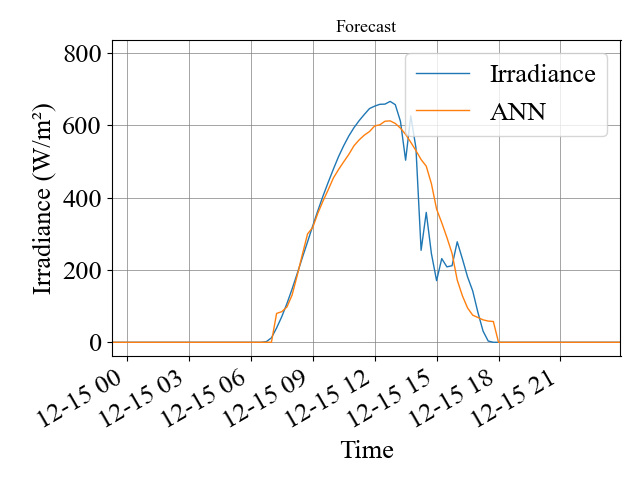}
        \caption{ANN irradiance forecast for December $15^{th}$.}
        \label{fig:ann_dec}
    \end{subfigure}
    \caption{Day forecasts of 15-minute irradiance values across the seasons from the ANN model.}
\end{figure*}
\begin{figure*}
    \vspace{0.42cm}
    \centering
    \captionsetup{justification=centering}
    \includegraphics[width=15cm,height=4.5cm,trim={0 0.9cm 0 1.4cm},clip]{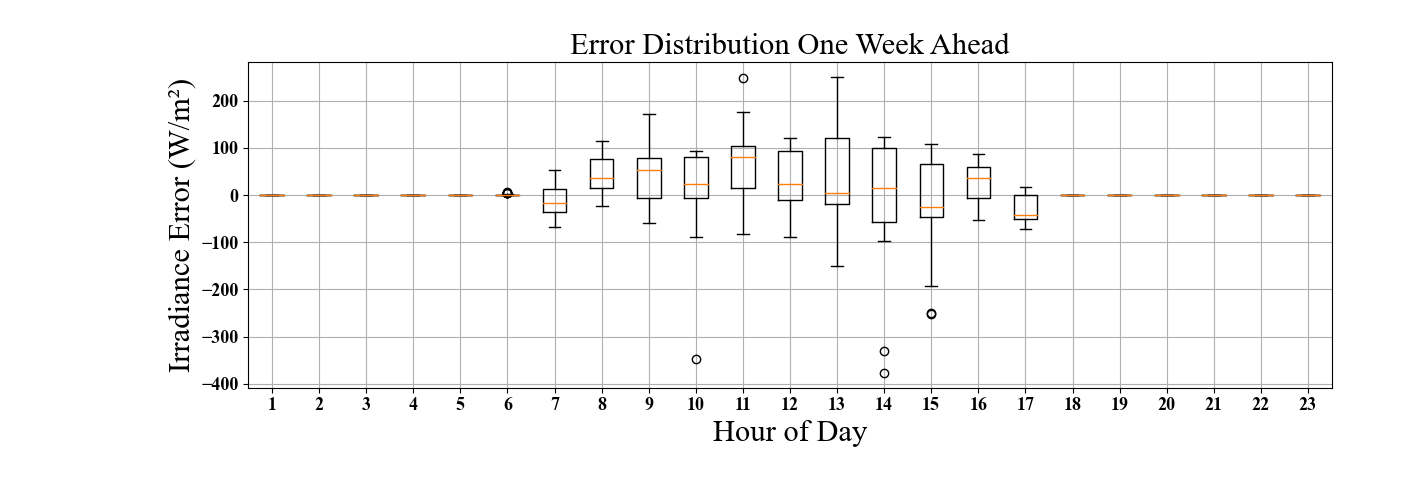}
    \caption{The hourly distribution of error in irradiance forecast of the ANN model.}
    \label{fig:error_ann}
\end{figure*}
\begin{figure*}[t]
    \captionsetup[subfigure]{justification=centering}
    \begin{subfigure}{\columnwidth}
        \centering
        \includegraphics[width=\columnwidth,trim={0 1.2cm 0 1cm},clip]{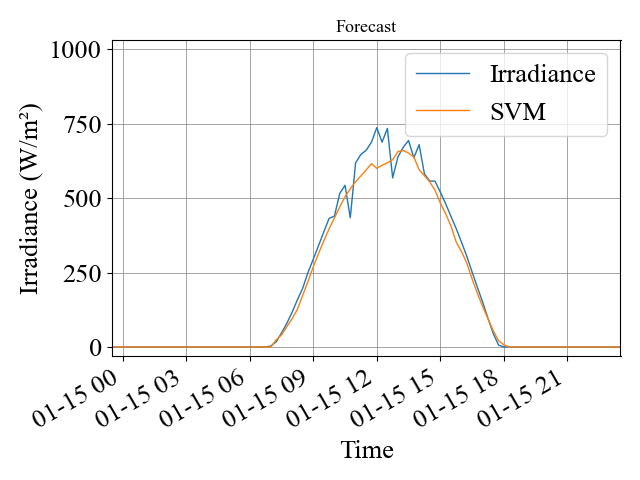}
        \caption{SVM irradiance forecast for January $15^{th}$.}
        \label{fig:SVM_hourly_jan}
    \end{subfigure} 
    \begin{subfigure}{\columnwidth}
        \centering
        \includegraphics[width=\columnwidth,trim={0 1.2cm 0 1cm},clip]{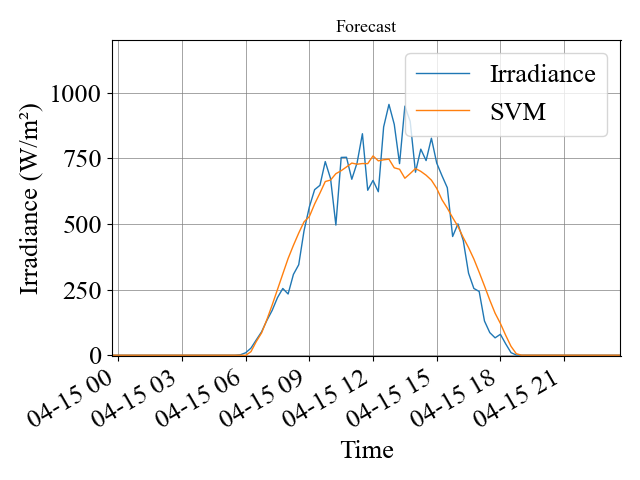}
        \caption{SVM irradiance forecast for April $15^{th}$.}
        \label{fig:SVM_hourly_apr}
    \end{subfigure}
    \begin{subfigure}{\columnwidth}
        \centering
        \includegraphics[width=\columnwidth,trim={0 1.2cm 0 1cm},clip]{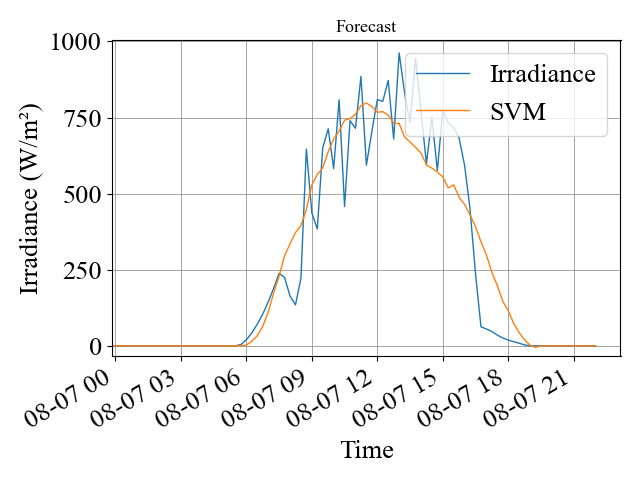}
        \caption{SVM irradiance forecast for August $7^{th}$.}
        \label{fig:SVM_hourly_aug}
        \end{subfigure}
    \hspace*{0.55cm}
    \begin{subfigure}{\columnwidth}
        \centering
        \includegraphics[width=\columnwidth,trim={0 1.2cm 0 1cm},clip]{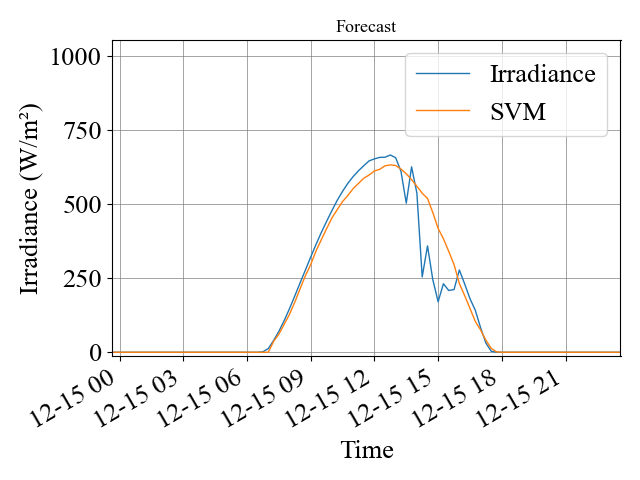}
        \caption{SVM irradiance forecast for December $15^{th}$.}
        \label{fig:SVM_hourly_dec}
    \end{subfigure}
    \caption{The forecast of intraday  irradiance.}
\end{figure*}
\begin{figure*}
    \vspace{0.42cm}
    \centering
    \captionsetup{justification=centering}
    \includegraphics[width=15cm,height=4.5cm,trim={0 0.9cm 0 1.4cm},clip]{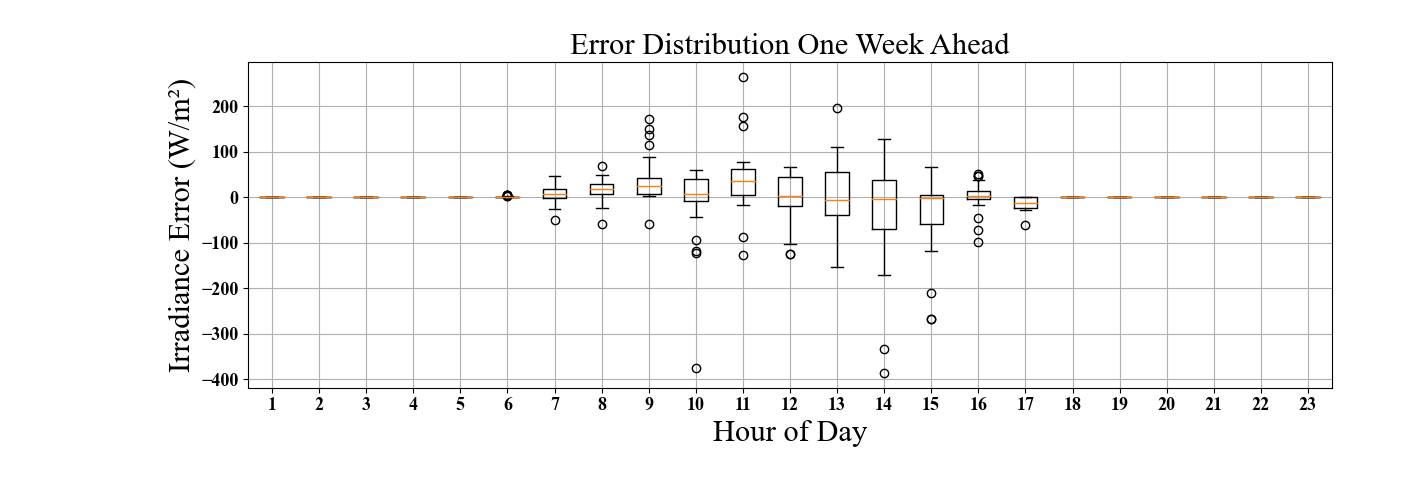}
    \caption{The hourly distribution of error in irradiance forecast of the SVM model.}
    \label{fig:SVM_error}
\end{figure*}
\begin{figure*}[h]
    \captionsetup[subfigure]{labelformat=empty,justification=centering}
    \begin{subfigure}{\columnwidth}
        \centering
        \includegraphics[width=\columnwidth,trim={0 1.2cm 0 1cm},clip]{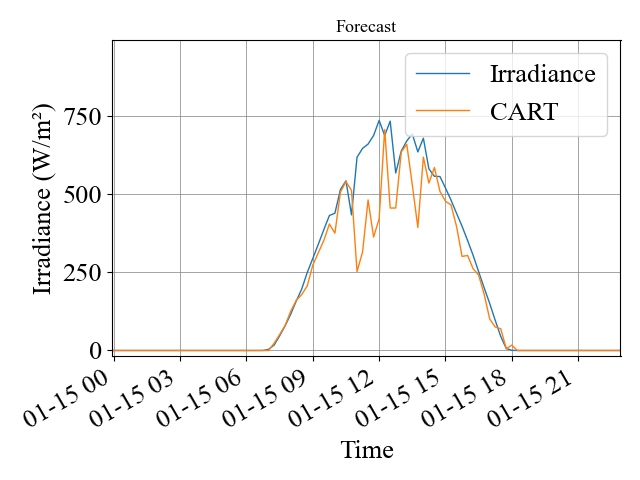}
        \caption{CART irradiance forecast for January $15^{th}$.}
        \label{fig:CART_jan}
    \end{subfigure} 
    \begin{subfigure}{\columnwidth}
        \centering
        \includegraphics[width=\columnwidth,trim={0 1.2cm 0 1cm},clip]{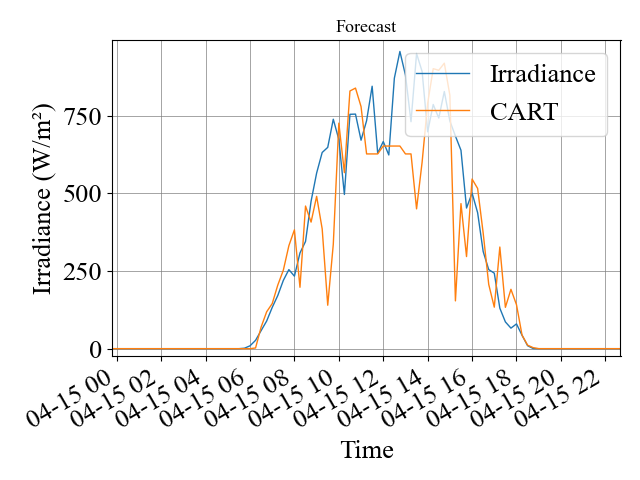}
        \caption{CART irradiance forecast for April $15^{th}$.}
        \label{fig:CART_apr}
    \end{subfigure}
    \begin{subfigure}{\columnwidth}
        \centering
        \includegraphics[width=\columnwidth,trim={0 1.2cm 0 1cm},clip]{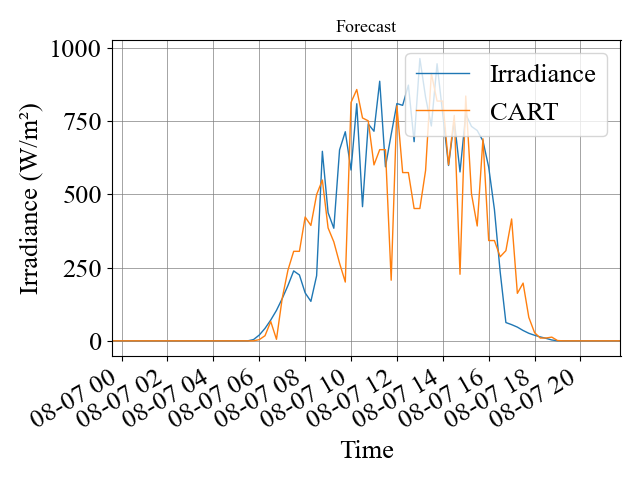}
        \caption{CART irradiance forecast for August $7^{th}$.}
        \label{fig:CART_aug}
        \end{subfigure}
    \hspace*{0.55cm}
    \begin{subfigure}{\columnwidth}
        \centering
        \includegraphics[width=\columnwidth,trim={0 1.2cm 0 1cm},clip]{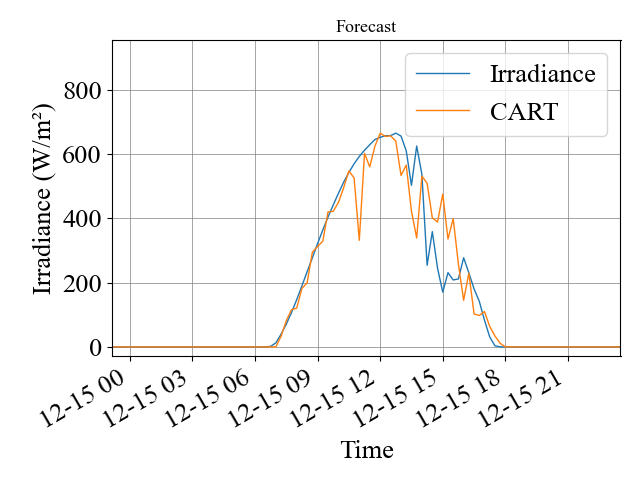}
        \caption{CART irradiance forecast for December $15^{th}$.}
        \label{fig:CART_dec}
    \end{subfigure}
    \caption{Day forecasts of 15-minute irradiance values across the seasons from the CART model.}
\end{figure*}
\begin{figure*}
    \centering
    \vspace{0.42cm}
    \captionsetup{justification=centering}
    \includegraphics[width=15cm,height=4.5cm,trim={0 0.9cm 0 1.4cm},clip]{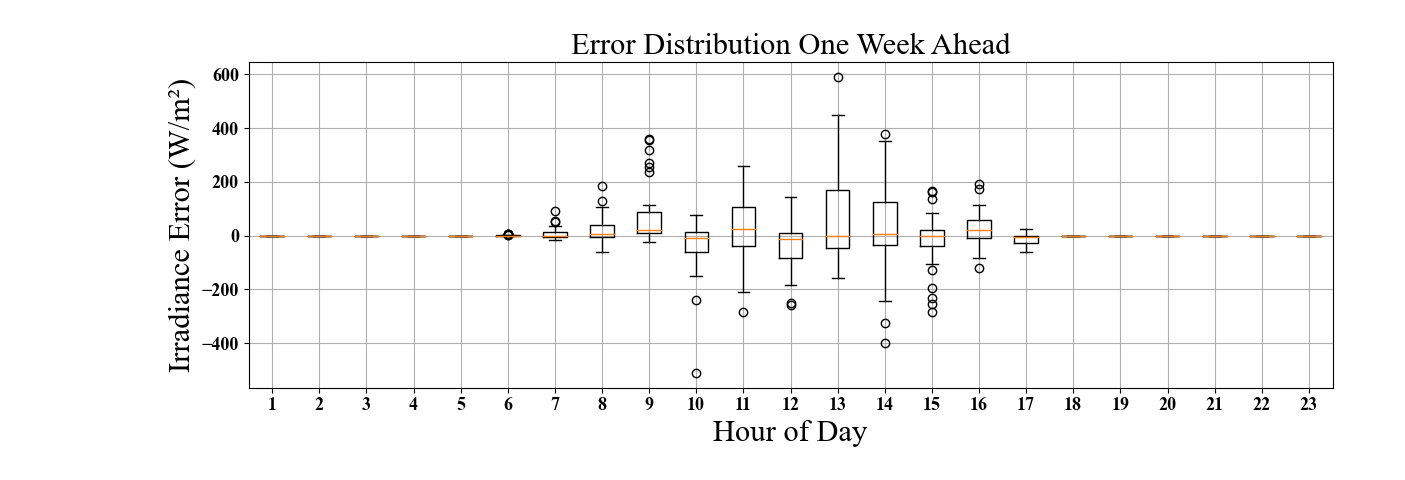}
    \caption{The hourly distribution of error in irradiance forecast of the CART model.}
    \label{fig:error_cart}
\end{figure*}
\begin{figure*}[h]
    \captionsetup[subfigure]{labelformat=empty,justification=centering}
    \begin{subfigure}{\columnwidth}
        \centering
        \includegraphics[width=\columnwidth,trim={0 1.2cm 0 1cm},clip]{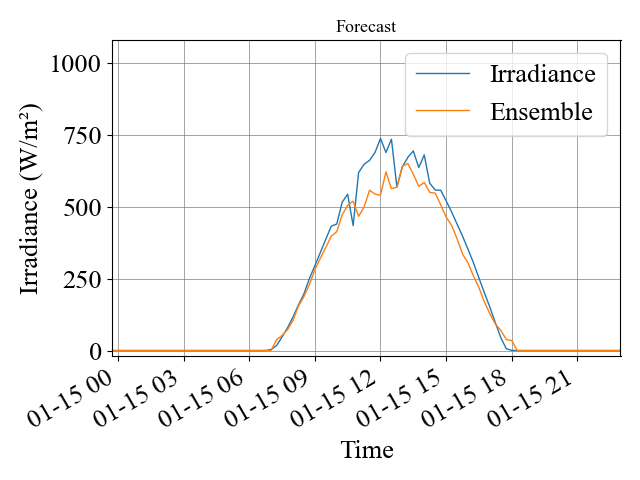}
        \caption{Ensemble irradiance forecast for January $15^{th}$.}
        \label{fig:ensemble_jan}
    \end{subfigure} 
    \begin{subfigure}{\columnwidth}
        \centering
        \includegraphics[width=\columnwidth,trim={0 1.2cm 0 1cm},clip]{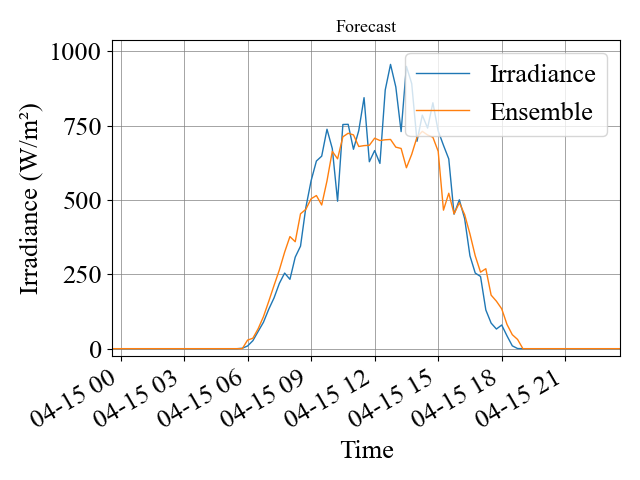}
        \caption{Ensemble irradiance forecast for April $15^{th}$.}
        \label{fig:ensemble_apr}
    \end{subfigure}
    \begin{subfigure}{\columnwidth}
        \centering
        \includegraphics[width=\columnwidth,trim={0 1.2cm 0 1cm},clip]{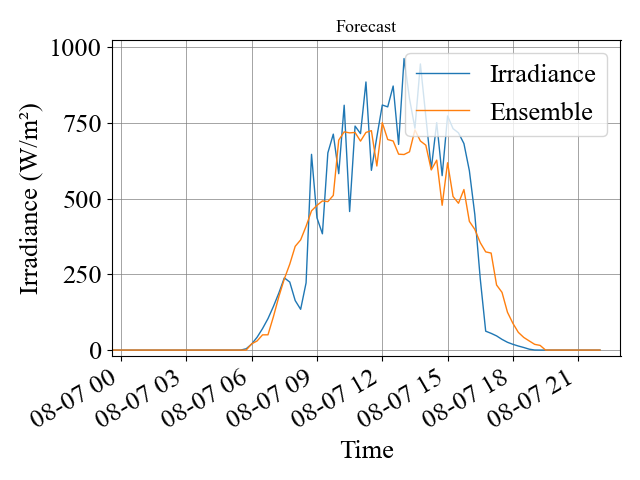}
       \caption{Ensemble irradiance forecast for August $7^{th}$.}
        \label{fig:ensemble_aug}
        \end{subfigure}
    \hspace*{0.55cm}
    \begin{subfigure}{\columnwidth}
        \centering
        \includegraphics[width=\columnwidth,trim={0 1.2cm 0 1cm},clip]{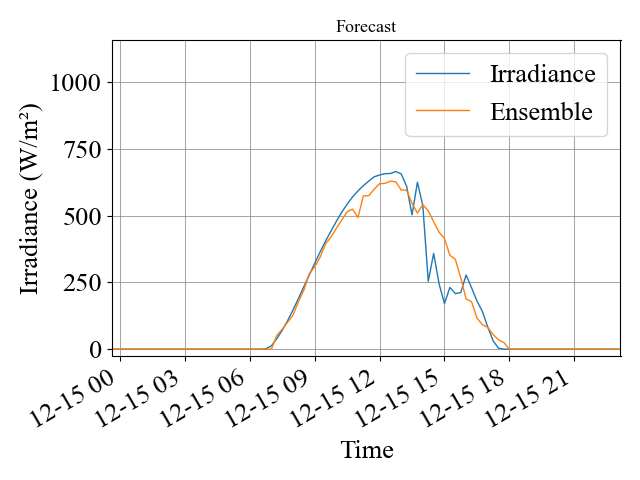}
       \caption{Ensemble irradiance forecast for December $15^{th}$.}
        \label{fig:ensemble_dec}
    \end{subfigure}
    \caption{Day forecasts of 15-minute irradiance values across the seasons from the ensemble model.}
\end{figure*}
\begin{figure*}
    \centering
    \vspace{0.42cm}
    \captionsetup{justification=centering}
    \includegraphics[width=15cm,height=4.5cm,trim={0 0.9cm 0 1.4cm},clip]{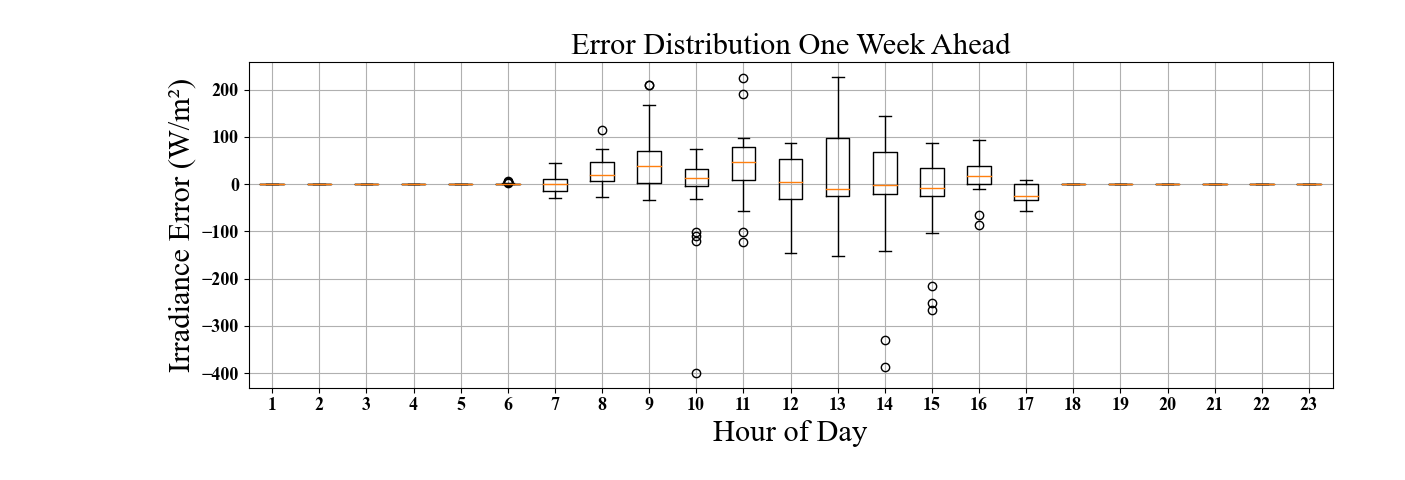}
    \caption{The hourly distribution of error in irradiance forecast of the ensemble model.}
    \label{fig:error_ensemble}
\end{figure*}
\begin{figure*}[t]
\centering
  \centering
  \includegraphics[width=16cm,height=8cm]{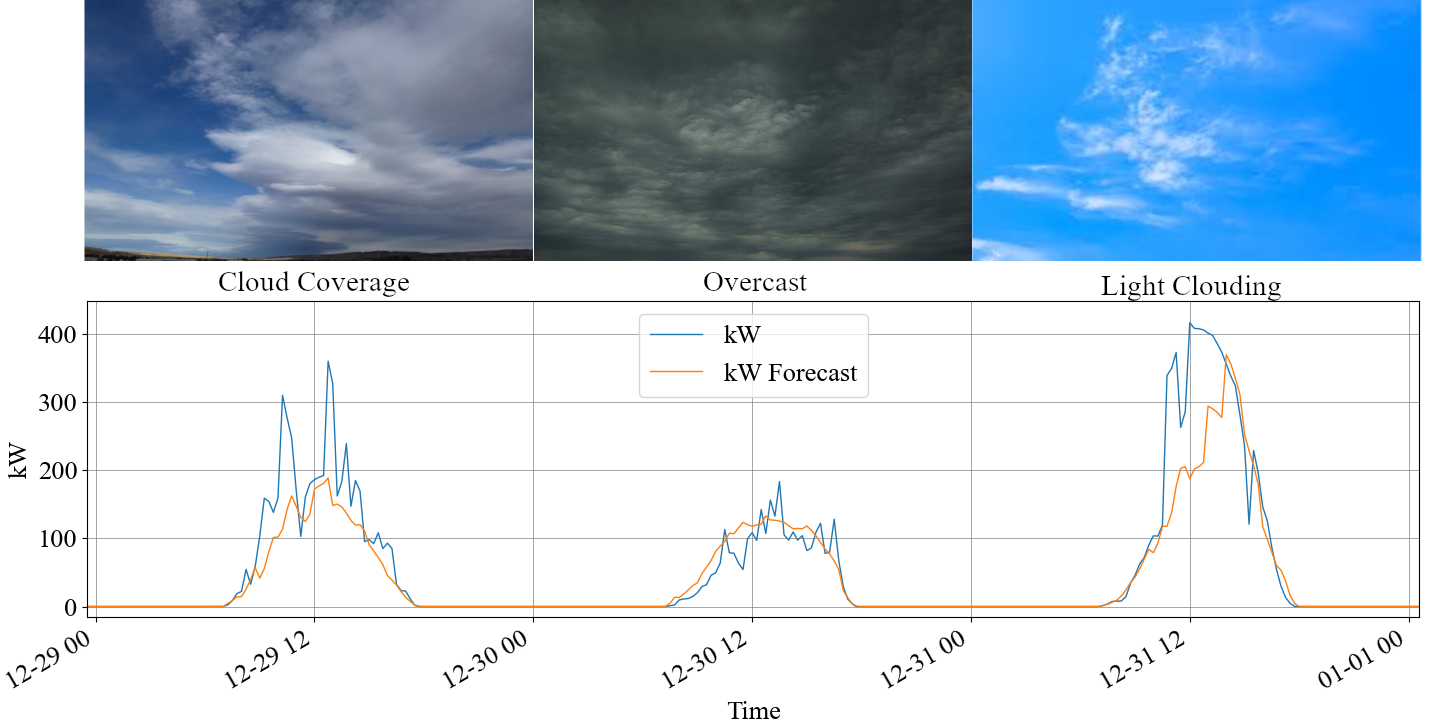}
  \caption{A set of three consecutive days of variable cloud coverage showing the ability of the forecaster to model the generation through high variability.}
  \label{fig:sub1}
\end{figure*}
\section{Results and Discussion}
\label{sec:results}
\textbf{Irradiation Forecast}
The irradiation, $\text{Whr/m}^2$, day ahead prediction is calculated based on the summation of daily forecasted irradiance captured in figures \ref{fig:ann_week_irradiation_daily}, \ref{fig:cart_week_irradiation_daily}, \ref{fig:svm_week_irradiation_daily}, \ref{fig:ensemble_week_irradiation_daily}. The forecasted irradiation summation per day is calculated for a week in May for all of the models. The irradiance in the test data set is predicted at a 15-minute time resolution and aggregated daily. The error metrics for hourly and daily forecasts are summarized at the end of the section.
\subsection{Varying Cloud Coverage}
Clear-day irradiance and generation are forecasted with the least error and increased cloud coverage is associated with the increased error in the forecast described in \cite{lorenz2009irradiance} and is further confirmed in the forecasts generated through this effort. The hourly error distributions in the clear sky and cloudy conditions, or cloud coverage \textit{cc}, are shown in figure \ref{fig:boxwhisk}. Despite the increased error in clouded conditions, the forecast is able to predict the generation across variable conditions. A set of days was chosen that highlights the forecast over light clouding, overcast, and cloudy days, showing that the forecast does capture the generation appropriately for the varying conditions. The error is calculated across a selection of light clouding <20\% \textit{cc}, cloudy 20\% < \textit{cc} < 50 \%, and overcast cc > 50\% days, and the error for these conditions confirms that the error will increase with the increase in the degree of \textit{cc}. In figure \ref{fig:clouderror} a selection of absolute error over morning hours from clear sky, cloudy, and overcast days show the error trend.  This demonstrates that, regardless of the low negative correlation between \textit{cc} and production, there is a strong effect from \textit{cc} variability on generation. 

\begin{table*}
\centering
    \captionsetup{justification=centering}
    \captionof{table}{Error metrics of 15-minute resolution intraday irradiance $\text{W}/\text{m}^2$ models.} 
    \label{tab:errorHourly}
\begin{tabular}{P{0.2\textwidth}|P{0.11\textwidth}|P{0.11\textwidth}|P{0.11\textwidth}|P{0.11\textwidth}}
\rowcolor[HTML]{EFEFEF} 
\hline
\textbf{Metric} & \textbf{SVM} & \textbf{CART} & \textbf{ANN} & \textbf{Ensemble} \\ \hline
MAE       & 35.804       & 53.006      & 38.9     &  37.7 \\ \hline
MSE       & 5941.952     & 12930.817    & 5622.596  &  6143.2 \\ \hline
RMSE      & 77.0841      & 113.714      & 78.378    &  78.379  \\ \hline
RRMSE     & 28\%         & 42\%      & 30\%    &  30\%  \\ \hline
$R^{2}$   & 0.87         & 0.74       & 0.88     & 0.88     \\ \hline
\end{tabular}
\end{table*}
\begin{table*}[t]
\centering
    \captionsetup{justification=centering}
    \captionof{table}{Error metrics of day ahead daily irradiation $\text{Whr}/\text{m}^2$ models.} 
    \label{tab:errorDaily}
\begin{tabular}{P{0.2\textwidth}|P{0.11\textwidth}|P{0.11\textwidth}|P{0.11\textwidth}|P{0.11\textwidth}}
\rowcolor[HTML]{EFEFEF} 
\hline
\textbf{Metric} & \textbf{SVM} & \textbf{CART} & \textbf{ANN} & \textbf{Ensemble} \\ \hline
MAE       &  16.763     & 17.694      & 16.142    &  15.531 \\ \hline
MSE       & 554.291      & 519.621    & 459.716   &  443.453 \\ \hline
RMSE      & 23.543      & 22.795      & 21.441    &  21.058  \\ \hline
RRMSE      & 14\%      & 14\%         & 13\%      &  13\%  \\ \hline
$R^{2}$      & 0.90      & 0.91       & 0.92      & 0.92     \\ \hline
\end{tabular}
\end{table*}
\textbf{Backcasting} is a use case for forecasting to replace missing data after major events or other interruptions. It is helpful when analyzing data sets for visibility and control of assets when there are gaps in the data. The loss of data may occur for many different reasons, software updates causing incompatibilities, power outages, interference in communications, and other disruptions. In 2017 Hurricane Irma disrupted data capture from the PV site in Miami. The advantage of the forecaster in the scenario of missing data is in the imputation of missing data during these disruptions. This is the advantage of back-casting through applying the forecast on historical data to compute an estimate for the performance of the system during the time when it was not visible. During such a major event it is likely that a gap in data will begin.
\subsection{Irradiance Forecast Across Season}
Intra-day forecast from the ANN model for January $15^{th}$ April $15^{th}$, August $7^{th}$, and December $15^{th}$ are shown  in figures \ref{fig:ann_jan}, \ref{fig:ann_apr}, \ref{fig:ann_aug}, and \ref{fig:ann_dec}. This shows the ability of the forecaster to accurately predict irradiance across the seasonal variations of spring, summer, fall, and winter throughout the year. The irradiance prediction error for each hour was calculated and is shown in the hourly box and whisker plot for the ANN in figure \ref{fig:error_ann}. The median error of the ANN does not exceed 100 W$m^2$ over days of varrying cloudy and clear conditions. The largest distribution of error is seen at 1 PM. The model has a bias towards over-forecasting in the hours from 8 AM till 4 PM, and a slight bias towards under-forecasting the hour after sunrise and the hour before sunset. \\

Intra-day forecast from the SVM model for January $15^{th}$ April $15^{th}$, August $7^{th}$, and December $15^{th}$ are shown  in figures \ref{fig:SVM_hourly_jan}, \ref{fig:SVM_hourly_apr}, \ref{fig:SVM_hourly_aug}, and \ref{fig:SVM_hourly_dec}. This shows the ability of the forecaster to accurately predict irradiance across the seasonal variations of spring, summer, fall, and winter throughout the year. The per-hour error irradiance prediction error was calculated and is shown in the hourly box and whisker plot for the SVM in figure \ref{fig:SVM_error}. The forecast is created from 5 AM in the morning and recalculated hourly and is used by the PV generation estimator to create 15-minute resolution KW forecast values. The forecast is generated up to seven days ahead and utilzies the regression models of module temperature and ideal irradiance along with the NWP cloud coverage data. The hourly error distributions, in figure \ref{fig:SVM_error}, show that the median of the error oscillates around zero suggesting no bias towards over or under-prediction of the irradiance in the SVM model. The error can be expected to increase during the highest production hours of the day near solar noon due to the intermittent spike pattern from interference with solar energy at this time of day. 

Intra-day forecast from the CART model for January $15^{th}$ April $15^{th}$, August $7^{th}$, and December $15^{th}$ are shown  in figures \ref{fig:CART_jan}, \ref{fig:CART_apr}, \ref{fig:CART_aug}, and \ref{fig:CART_dec}. This shows the ability of the forecaster to accurately predict irradiance across the seasonal variations of spring, summer, fall, and winter throughout the year. The irradiance prediction error for each hour was calculated and is shown in the hourly box and whisker plot for the CART in figure \ref{fig:error_cart}. The median error of the CART is similar to that of the SVM model and suggests that the CART model has a low or no bias towards under or over-forecasting. Similarly to the SVM model the CART model is generating more outlier data points in comparison with the ANN model.

Intra-day forecast from the ensemble model for January $15^{th}$ April $15^{th}$, August $7^{th}$, and December $15^{th}$ are shown  in figures \ref{fig:ensemble_jan}, \ref{fig:ensemble_apr}, \ref{fig:ensemble_aug}, and \ref{fig:ensemble_dec}. This shows the ability of the forecaster to accurately predict irradiance across the seasonal variations of spring, summer, fall, and winter throughout the year. The irradiance prediction error for each hour was calculated and is shown in the hourly box and whisker plot for the CART in figure \ref{fig:error_ensemble}. The median error of the ensemble model is similar to that of the SVM model and suggests that the ensemble model has a low or no bias towards under or over-forecasting. The ensemble model is also producing less outliers than the SVM and CART models.
\\

The seventh day of August is chosen due to gaps in the actual data. It can be seen that the models have a very low bias towards over or under-predicting irradiance. The mean error across the hourly box and whisker charts of hourly error distribution is oscillating around zero. The exception is the ANN model which tends to over estimate irradiance.

\begin{figure*}[h]
    \captionsetup[subfigure]{labelformat=empty,justification=centering}
    \begin{subfigure}{0.5\columnwidth}
        \centering
        \includegraphics[width=\columnwidth,trim={0 1.2cm 0 1cm},clip]{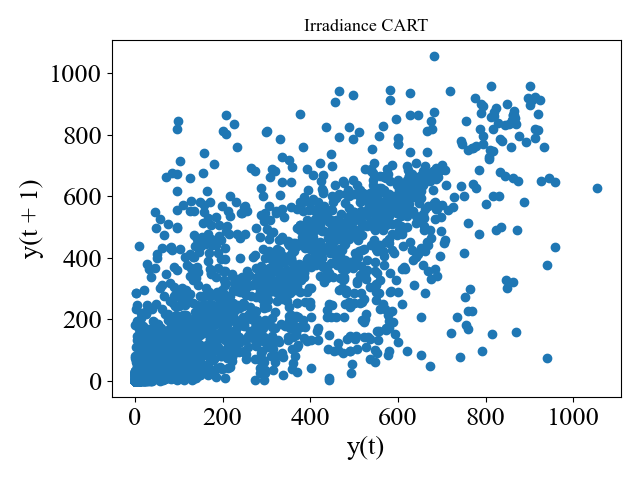}
        \caption{The CART Irradiance forecast lag plot.}
        \label{fig:cartlag}
    \end{subfigure} 
    \begin{subfigure}{0.5\columnwidth}
        \centering
        \includegraphics[width=\columnwidth,trim={0 1.2cm 0 1cm},clip]{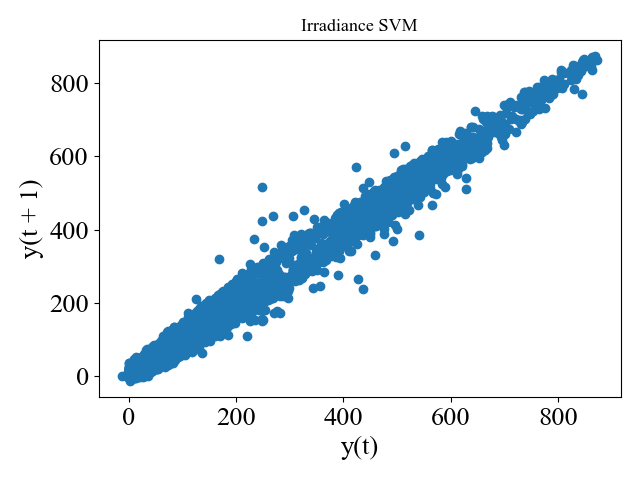}
        \caption{The SVM Irradiance forecast lag plot.}
        \label{fig:svmlag}
    \end{subfigure}
    \begin{subfigure}{0.5\columnwidth}
        \centering
        \includegraphics[width=\columnwidth,trim={0 1.2cm 0 1cm},clip]{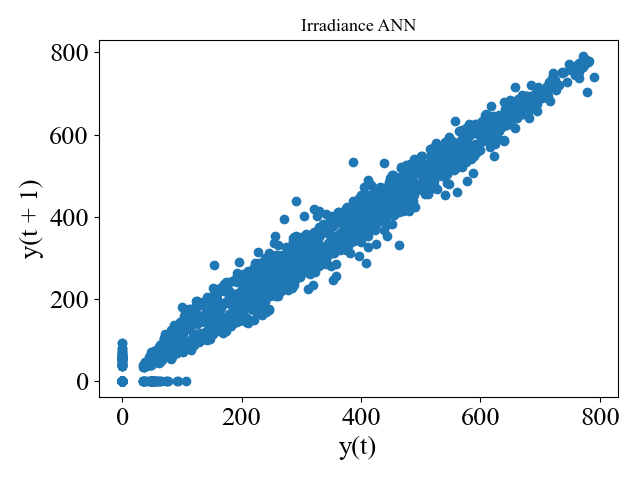}
        \caption{The ANN Irradiance forecast lag plot.}
        \label{fig:annlag}
        \end{subfigure}
    \begin{subfigure}{0.5\columnwidth}
        \centering
        \includegraphics[width=\columnwidth,trim={0 1.2cm 0 1cm},clip]{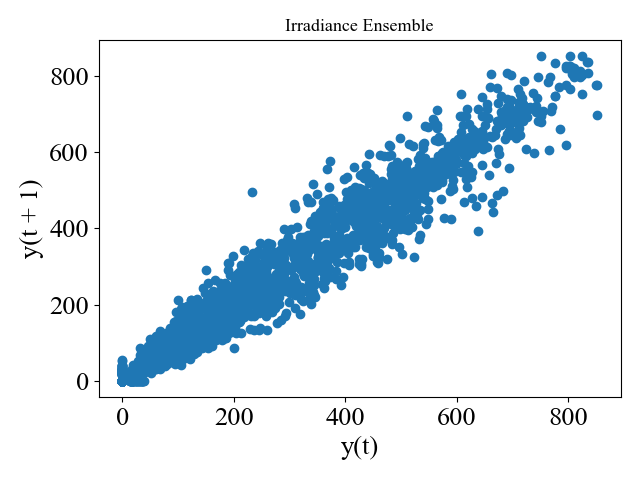}
        \caption{The Ensemble Irradiance forecast lag plot.}
        \label{fig:ensemblelag}
    \end{subfigure}
    \caption{Lag plots (lag=1) of actual irradiance, and the forecast test data from the CART, SVM, ANN, and Ensemble models.}
\end{figure*}
\begin{figure}{6cm}
    \centering
    \includegraphics[width=\columnwidth,trim={0 1.2cm 0 1cm},clip]{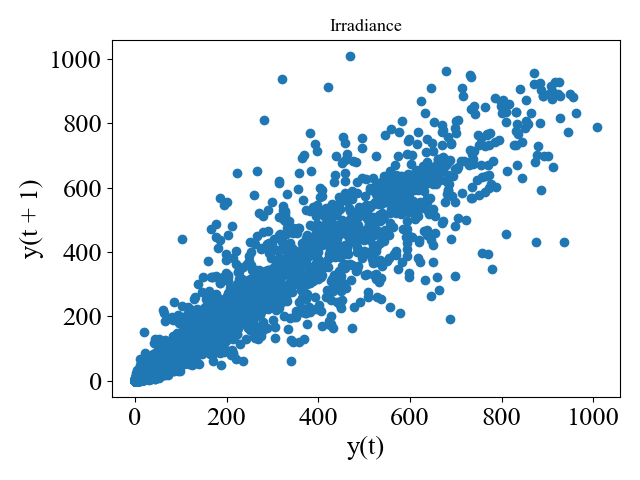}
    \caption{The lag plot from actual Irradiance recordings.}
    \label{fig:ensemblelag}
\end{figure}




\textbf{Lag Plots:}
The lag plots show that actual irradiance increases variability from 200 $\text{Wm}^2$ through 1000 $\text{Wm}^2$. The CART model has a high level of variation throughout 0 $\text{Wm}^2$ through 1000 $\text{Wm}^2$. The ensemble model more closely models the variability of the actual generation and it shows increased variability from 200 $\text{Wm}^2$ through 1000 $\text{Wm}^2$. The ensemble model performs best in terms of matching the variability pattern of the actual irradiance. A time-dependent mixture density network or LSTM model could capture the variability pattern of actual irradiance \cite{Sarochar}. The ANN and SVM models fail to capture the variation.

\textbf{Use cases of a PV forecast:}
There are many reasons to utilize forecasts the first being visibility of assets performance and planning for use of generated energy. The application of forecasting can be used in addressing the duck-curve phenomena that have been observed in high penetration PV in the California energy market. The forecast can be used in applications for asset owners of microgrids in which energy storage is available. Energy arbitrage is a potential use case for the forecast of PV when coordinated with the storage asset the energy can be used at other times of the day when the cost of grid-imported energy is higher. Peak shaving is yet another potential application by storing peak PV generation in in battery systems for a later discharge of the stored energy. The ability to forecast PV can enhance access to flexible generation for the grid and will play a role in the energy market especially as higher penetration of PV into the energy markets become more common over time. 


\section{Conclusion}
\label{sec:conclusion}

This paper makes its contribution by showing a forecaster of PV array output power. The holistic approach was trained and tested on a 1.4 MW DC PV power plant in Florida. The  forecasting uses solar variables as inputs and predicts the irradiance, from this an estimator creates the final expected power generation. Using linear regression models to project solar module temperature, and for short-term forecasting, and relying on the NWP and other data for longer intra-day and weekly forecasting. The pattern of irradiance shows increasing variability in the hours around solar noon, and an ensemble model can more accurately capture this dynamic by combining the outputs of various models. The potential of LSTM and MDN models to capture this time-dependent pattern of near-solar noon variability is a direction for further research. Hyperparameter tuning such as adjustment of the SVM parameters can improve the model output for this problem. Although matching the variability pattern of the actual irradiance may be possible with CART, it will be important not to have an inaccurate spike pattern in the intra-day forecast of applications that use short-term forecasts. The ensemble model for hourly irradiance forecast is a fair performance model and the daily forecast ensemble model performs well. 
\section*{NOMENCLATURE}
\begin{table}[h]
\normalsize
\begin{tabular}{ll}
ANN         & Artificial Neural Network      \\
R           & Irradiance                     \\
R0          & Ideal Irradiance               \\
RMSE        & Root Mean Square Error         \\
MAE         & Mean Absolute Error            \\
MAPE        & Mean Absolute Percentage Error \\
MSE         & Mean Square Error              \\
API         & Application Program Interface  \\
NWP         & Numerical Weather Prediction \\
CART        & Classification and Regression Tree \\
SVM         & Support Vector Machine           \\
JD          & Julian Date                   \\
$\phi s$    & Solar Azimuth                  \\
$d$         & Distance of sun from earth     \\
$\delta$    & Solar declination              \\
$\beta_{N}$ & Irradiance angle of incidence \\
$\beta$ & Altitude Angle of the sun\\
\end{tabular}
\end{table}
\section{Acknowledgement}
The authors would like to acknowledge the Department of Electrical and Computer Engineering (ECE) at the FIU College of Engineering for supporting this work.

\bibliographystyle{IEEEtran}
\bibliography{ref.bib}

\EOD
\end{document}